\documentclass[%
 reprint,
superscriptaddress,
 amsmath,amssymb,
 aps,
prb,
]{revtex4-2}

\usepackage{graphicx}
\usepackage{dcolumn}
\usepackage{bm}

\usepackage{color}
\usepackage{textcomp}
\usepackage{mathcomp}
\usepackage{multirow}

\begin{document}


\title{\textbf {
Anisotropic spin fluctuations in the triangular Kondo lattice  compound CePtAl$_4$Ge$_2$ probed by site-selective $^{27}$Al NMR
} 
}%

\author{H. Sakai}
\email{sakai.hironori@jaea.go.jp}
\affiliation{Advanced Science Research Center, Japan Atomic Energy Agency, Tokai, Ibaraki 319-1195, Japan}

\author{S. Shin}
\affiliation{PSI Center for Neutron and Muon Sciences, Paul Scherrer Institut, 5232
Villigen PSI, Switzerland.}
\affiliation{Center for Quantum Materials and Superconductivity (CQMS) and Department of
Physics, Sungkyunkwan University, Suwon 16419, South Korea.}
\altaffiliation{Current affiliation: J\"ulich Centre for Neutron Science (JCNS) at the Heinz Maier-Leibnitz Zentrum (MLZ), Forschungszentrum J\"ulich GmbH, Lichtenbergstra{\ss}e 1, 85747, Garching, Germany.}

\author{S. Kambe}
\affiliation{Advanced Science Research Center, Japan Atomic Energy Agency, Tokai, Ibaraki 319-1195, Japan}

\author{Y. Tokunaga}
\affiliation{Advanced Science Research Center, Japan Atomic Energy Agency, Tokai, Ibaraki 319-1195, Japan}

\author{H. Harima}
\affiliation{Graduate School of Science, Kobe University, Kobe 657-8501, Japan}

\author{E. Pomjakushina}
\affiliation{PSI Center for Neutron and Muon Sciences, Paul Scherrer Institut, 5232
Villigen PSI, Switzerland.}

\author{T. Park}
\affiliation{Center for Quantum Materials and Superconductivity (CQMS) and Department of
Physics, Sungkyunkwan University, Suwon 16419, South Korea.}

\date{October 24, 2025}

\begin{abstract}
A site-selective $^{27}$Al nuclear magnetic resonance (NMR) study is carried out on the Kondo lattice compound CePtAl$_4$Ge$_2$, which crystallizes in a rhombohedral lattice with quasi-two-dimensional Ce layers forming a triangular lattice network.
Two inequivalent Al sites, Al(1) and Al(2), are unambiguously assigned by comparing measured nuclear quadrupole parameters with electric field gradients obtained from electronic structure calculations.
Knight shift analysis yields distinct hyperfine coupling constants, revealing that they arise predominantly from Ruderman-Kittel-Kasuya-Yosida (RKKY)-type transferred hyperfine fields through conduction electrons.
Spin-lattice relaxation measurements reveal pronounced anisotropic spin fluctuations, and comparison of the relaxation rates between the two Al sites clarifies the momentum-space structure of these fluctuations.
At low magnetic fields, $(T_1T)^{-1}$ is strongly enhanced on cooling toward the N\'eel temperature, indicating the growth of in-plane antiferromagnetic correlations in the paramagnetic state.
\end{abstract}

\maketitle


\section{\label{sec:introduction}Introduction}

Frustrated magnetism in $f$-electron Kondo systems offers fertile ground for unconventional spin states, where the competition between Ruderman–Kittel–Kasuya–Yosida (RKKY) interactions and Kondo screening is strongly shaped by lattice geometry \cite{Si2006Global-magnetic, Vojta2008From-itinerant-, Coleman2010Frustration-and, Batista2016Frustration-and}.
Indeed, in Kondo lattice compounds where rare-earth or actinide ions form geometrically frustrated networks, such frustration can stabilize partial magnetic order and complex spin textures that are absent in conventional Kondo lattice compounds.
Yet, a detailed microscopic understanding of their spin dynamics remains limited, particularly regarding how geometrical frustration influences low-energy spin fluctuations in the presence of anisotropic interactions.

Representative examples such as CePdAl \cite{Donni1996Geometrically-f, Oyamada2008Ordering-mechan, Zhao2016Temperature-fie, Lucas2017Entropy-Evoluti} and UNi$_4$B \cite{Mentink1994Magnetic-Orderi, Mentink1995Magnetization-a, Movshovich1999Second-Low-Temp} feature quasi-kagome or distorted triangular lattice networks.
In these compounds, partial magnetic order emerges as a consequence of selective Kondo screening within geometrically constrained networks. Such behavior illustrates how magnetic frustration can suppress conventional long-range order and give rise to exotic spin textures in $f$-electron systems.

In contrast, trigonal CePtAl$_4$Ge$_2$ hosts a triangular lattice network of Ce ions stacked along the $c$ axis \cite{Shin2018Synthesis-and-c}, providing a structurally simpler platform to explore frustrated Kondo magnetism without site disorder.
Neutron diffraction has revealed a longitudinal spin-amplitude-modulated magnetic structure in this compound below $T_{\rm N}=2.3$~K \cite{Shin2020Magnetic-struct, Shin2023Triple-sinusoid_ArXiv}, positioning it as a rare system where frustration, Kondo coherence, and magnetic anisotropy interplay in a clean setting.
Compositional relatives such as the tetragonal Ce$M$Al$_4$Si$_2$ family ($M$ = Rh, Ir, Pt), where Ce ions form square lattices, lack geometrical frustration but exhibit a range of magnetic ground states from antiferromagnetic to ferromagnetic \cite{GhimireNJ:JPCM27:2015, Maurya2016Kondo-Lattice-a}.
Among them, CeRhAl$_4$Si$_2$ has been studied using $^{27}$Al nuclear magnetic resonance (NMR), revealing anisotropic spin fluctuations associated with Fermi surface nesting \cite{SakaiH:PRB93:2016}, and illustrating the power of site-selective NMR in probing low-energy spin dynamics.

In this work, a $^{27}$Al NMR study on single-crystalline CePtAl$_4$Ge$_2$ is presented to investigate its local magnetic properties and spin fluctuations under geometrical frustration.
By assigning the two inequivalent Al sites using electric field gradients (EFGs) obtained from electronic structure calculations, site-resolved Knight shift and spin-lattice relaxation measurements are carried out.
The results reveal highly anisotropic, predominantly in-plane spin fluctuations and their suppression under magnetic fields, offering microscopic insights into frustration-driven spin dynamics in a triangular $f$-electron system.

\section{\label{sec:experimental}Experimental Details}

Single crystals of CePtAl$_4$Ge$_2$ were grown using a self-flux method with an Al$_{0.73}$Ge$_{0.27}$ mixture, as described in Ref.~\onlinecite{Shin2018Synthesis-and-c}.
X-ray diffraction confirmed the rhombohedral lattice and single-phase quality of the crystals.
The chemical composition was verified using energy-dispersive x-ray spectroscopy.
For NMR measurements, a single crystal was mounted on a glass slip and tightly wound with copper wire to form a radio-frequency (rf) coil.
This coil was placed in a standard $^4$He cryostat equipped with a superconducting magnet and a variable-temperature insert.
A dual-axis goniometer enabled rotation of the sample with respect to the external magnetic field.
For measurements at temperatures between approximately 100 mK and 2 K, a $^3$He–$^4$He dilution refrigerator was used without the goniometer.
$^{27}$Al NMR (nuclear spin $I = 5/2$) spectra were acquired using a spin-echo technique with a phase-coherent pulsed NMR spectrometer.
The spectra were constructed using Fourier transforms of the spin-echo signals collected over a range of carrier frequencies.
The nuclear quadrupole frequency is defined as $\nu_{\rm Q} \equiv |3e^2qQ / {2I(2I-1)h}|$,
where $eQ$ is the nuclear quadrupole moment and $eq \equiv V_{ZZ}$ is the largest principal component of the EFG tensor.
The principal components of the EFG ($V_{ii}$) satisfy $|V_{XX}| \le |V_{YY}| \le |V_{ZZ}|$ and $V_{XX} + V_{YY} + V_{ZZ} = 0$.
The asymmetry parameter is defined as $\eta \equiv |V_{YY} - V_{XX}| / |V_{ZZ}|$.
We used the nuclear quadrupole moment for $^{27}$Al of $^{27}Q = 0.1466 \times 10^{-28}$ m$^2$ \cite{Pyykko2008Year-2008-nucle} and the nuclear gyromagnetic ratio $^{27}\gamma_{\rm n}/2\pi = 11.094$ MHz/T.
The nuclear spin-lattice relaxation time $T_1$ was determined using the inversion-recovery method.
Magnetization recovery curves were well fitted by single-component relaxation functions for both the second satellite transitions:
$\{ M(\infty) - M(t) \} / M(\infty) = \frac{1}{35}e^{-t/T_1} + \frac{3}{14}e^{-3t/T_1} + \frac{2}{5}e^{-6t/T_1} + \frac{2}{7}e^{-10t/T_1} + \frac{1}{14}e^{-15t/T_1}$, corresponding to the expected relaxation functions for spin $I = 5/2$ nuclei under magnetic relaxation.

\section{\label{sec:results}Results and discussions}

\subsection{$^{27}$Al NMR spectra and site-specific assignments}
\begin{figure}[hbt]
\includegraphics[width=8.5cm]{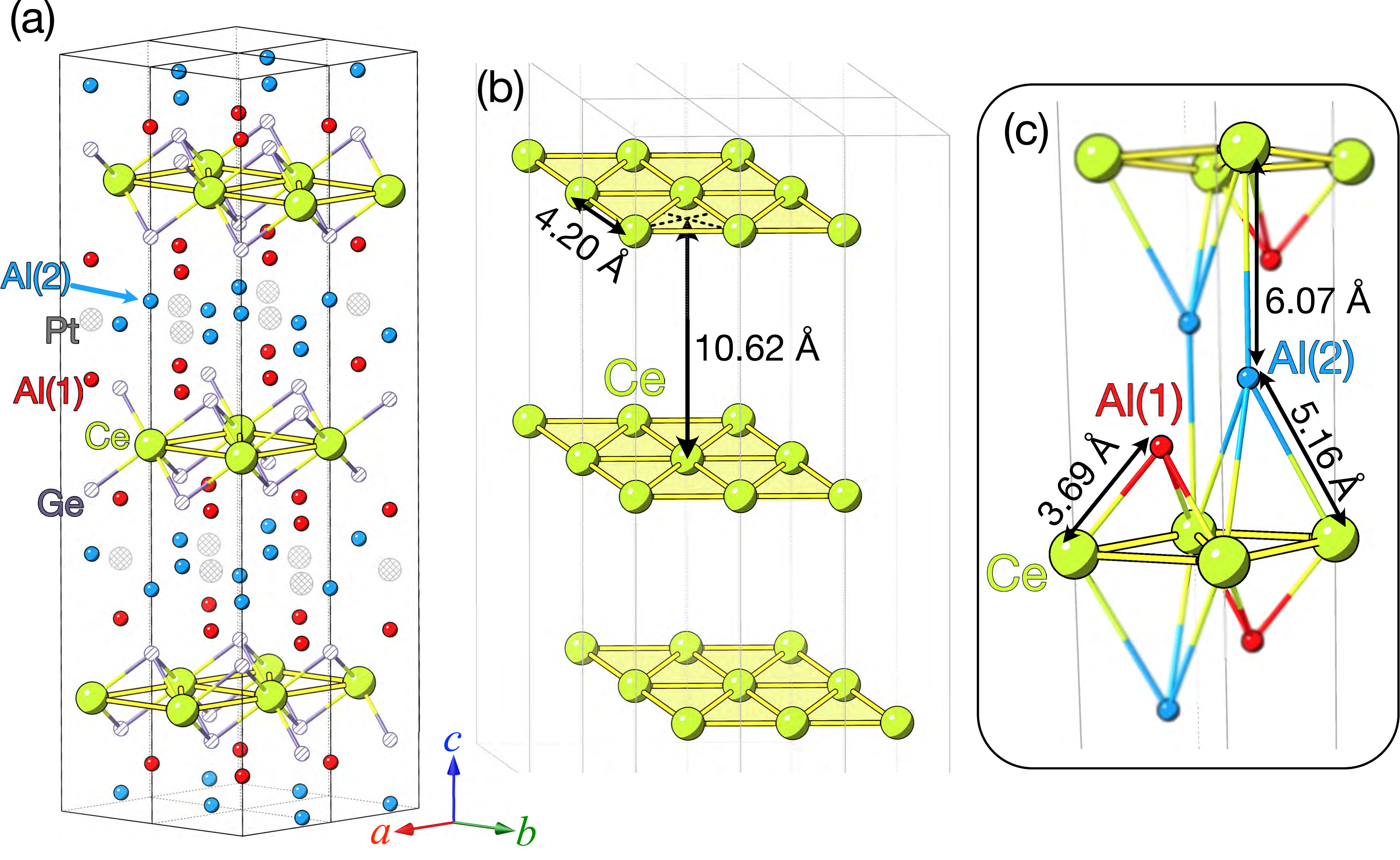}
\caption{\label{fig:CrystalStructure}
(a) Crystal structure of CePtAl$_4$Ge$_2$, consisting of alternating Ce layers and PtAl$_4$Ge$_2$ slabs stacked along the $c$ axis. (b) Extracted Ce sublattice highlighting the two-dimensional triangular lattice network. (c) Local environments of the two inequivalent Al sites, Al(1) and Al(2). Pt and Ge atoms are omitted for clarity.
}
\end{figure}

To interpret the $^{27}$Al NMR spectra in CePtAl$_4$Ge$_2$, it is essential to understand its crystal structure and the local environments of the Al sites, which determine the hyperfine and quadrupolar interactions.
CePtAl$_4$Ge$_2$ crystallizes in the YNiAl$_4$Ge$_2$-type structure with a rhombohedral lattice (space group $R\bar{3}m$, No.166), as illustrated in Fig.\ref{fig:CrystalStructure}(a) \cite{Shin2018Synthesis-and-c}.
The structure consists of alternating Ce layers and PtAl$_4$Ge$_2$ slabs stacked along the $c$ axis, with lattice parameters $a = 4.1995$\AA\ and $c = 31.851$\AA.
The Ce atoms form a well-separated triangular-lattice network, giving rise to a two-dimensional frustrated geometry [Fig.~\ref{fig:CrystalStructure}(b)].
Two inequivalent Al sites, Al(1) and Al(2), both occupy Wyckoff $6c$ positions with local $3m$ symmetry, as shown in Fig.\ref{fig:CrystalStructure}(c).

\begin{figure}[htb]
\includegraphics[width=8.5cm]{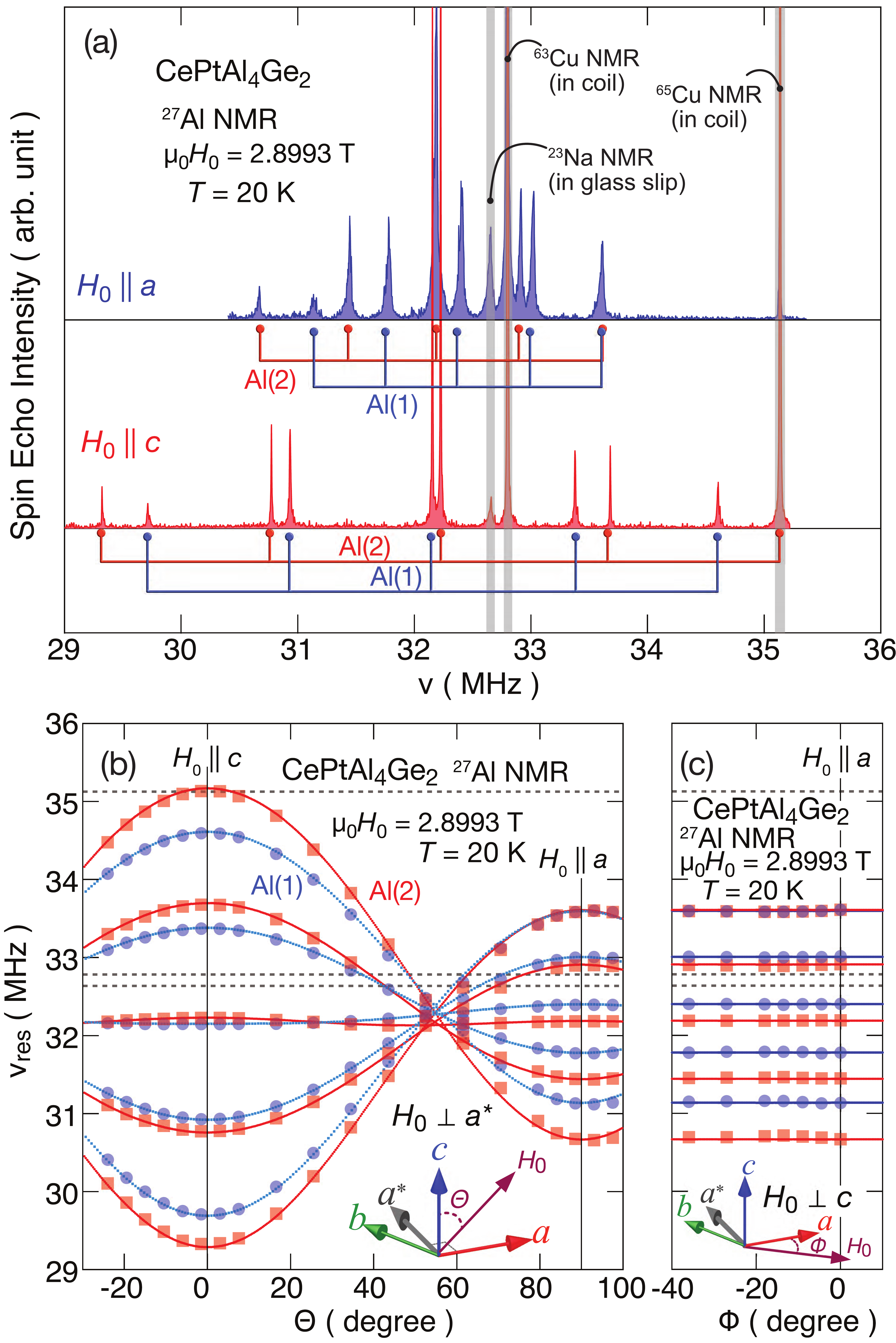}
\caption{\label{fig:NMRspectra}
$^{27}$Al NMR spectra of CePtAl$_4$Ge$_2$ at 20 K under a magnetic field of $\mu_0H_0 = 2.8993$ T.
(a) Field-swept spectra measured with $H_0 \parallel a$ and $H_0 \parallel c$. The two sets of lines are assigned to the Al(1) and Al(2) sites based on comparisons with electric field gradient parameters estimated from band structure calculations. Gray shaded regions indicate spectral blind spots due to $^{63,65}$Cu NMR signals from the NMR coil and $^{23}$Na NMR signals from the glass strip used to mount the crystal.
(b) Polar-angle ($\Theta$) dependence of the resonance frequencies for $H_0$ rotated within the $ac$ plane. Blue circles and red squares represent experimental data, while the bold blue dashed and red solid lines are simulations for the Al(1) and Al(2) sites, respectively, obtained by exact diagonalization of the nuclear spin Hamiltonian. Horizontal dashed lines mark the positions of extrinsic $^{63,65}$Cu and $^{23}$Na NMR signals.
(c) Azimuthal-angle ($\Phi$) dependence of the resonance frequencies for $H_0$ rotated within the $ab$ plane ($H_0 \perp c$).
}
\end{figure}

Figure~\ref{fig:NMRspectra}(a) shows the $^{27}$Al NMR spectra recorded at 20~K under $\mu_0H_0 = 2.8993$~T for both $H_0 \parallel a$ and $H_0 \parallel c$.
Each spectrum contains five quadrupole-split lines from the $I=5/2$ nuclei at the Al(1) and Al(2) sites.
No additional splitting is observed upon rotation, indicating that each site maintains equivalent local symmetry under the field.

Both Al sites share the same axial $3m$ symmetry, ensuring that the EFG tensor is axially symmetric with asymmetry parameter $\eta = 0$.
This is consistent with the observed angular dependence: satellite splitting is maximized for $H_0 \parallel c$, and no in-plane anisotropy is seen under azimuthal rotation, as shown in Fig.~\ref{fig:NMRspectra}(c).
These features confirm that the EFG principal axis is aligned along the $c$ axis at both sites.

The observed spectra are reproduced by numerically diagonalizing the nuclear spin Hamiltonian
$\mathcal{H} = \mathcal{H}_{\rm Z} + \mathcal{H}_{\rm Q}$, where 
\begin{equation}
\mathcal{H}_{\rm Z} = \gamma_{\rm n} \hbar \left\{1 + K(\Theta)\right\} \bm{I} \cdot \bm{H}_0,
\end{equation}
and
\begin{equation}
\mathcal{H}_{\rm Q} = \frac{h \nu_{\rm Q}}{6} \left(3 I_z^2 - I(I+1)\right).
\end{equation}
Assuming axial anisotropy in the Knight shift, the angle-dependent shift is modeled as
$K(\Theta) = K_c \cos^2\Theta + K_a \sin^2\Theta$,
where $K_a$ and $K_c$ denote the Knight shifts along the $a$ and $c$ axes.
The parameters $K_a$, $K_c$, and $\nu_{\rm Q}$ are optimized to fit the observed angular dependence of the resonance frequencies.

The best-fit parameters are $\nu_{\rm Q} = 1.23 \pm 0.01$~MHz, $K_a = 0.665 \pm 0.001$\%, and $K_c = -0.04 \pm 0.01$\% for the Al(1) site, and $\nu_{\rm Q} = 1.47 \pm 0.01$~MHz, $K_a = -0.031 \pm 0.001$\%, and $K_c = 0.20 \pm 0.01$\% for the Al(2) site.
Since both sites share the same local symmetry, these values alone do not uniquely determine the site assignment.
To resolve this ambiguity, nuclear quadrupole frequencies are compared with calculated values $\nu_{\rm Q}^{\rm calc}$.
The $\nu_{\rm Q}^{\rm calc}$ is obtained from the electronic structure calculations
using the full-potential linearized augmented plane
wave (FLAPW) method based on the local density approximation
(LDA), with the procedure of Blaha {\it et al.}~\cite{Blaha_rutile_EFGcalc}.
This approach has proven reliable in various Ce-based intermetallics, including CeRu$_2$Al$_{10}$ \cite{Matsumura2009Novel-Phase-Tra}, CeRu$_2$Al$_2$B~\cite{MatsunoH:JPSJ81:2012}, CeRu$_2$Ga$_2$B~\cite{SakaiH:PRB86:2012}, and CeRhAl$_4$Si$_2$~\cite{SakaiH:PRB93:2016}.
For LaPtAl$_4$Ge$_2$ and CePtAl$_4$Ge$_2$, the calculated $\nu_{\rm Q}^{\rm calc}$ values at the Al(1) and Al(2) sites are summarized in Table~\ref{tab:NQR_hyperfine}.
For both LaPtAl$_4$Ge$_2$ and CePtAl$_4$Ge$_2$, the $\nu_{\rm Q}^{\rm calc}$ at the Al(1) site is consistently smaller than that at the Al(2) site.
This lower $\nu_{\rm Q}^{\rm calc}$ value in each compound is in very good agreement with the experimentally observed $\nu_{\rm Q}^{\rm exp} = 1.23$~MHz, thereby confirming the assignment of the lower $\nu_{\rm Q}^{\rm exp}$ signal to the Al(1) site.
The quadrupole frequency shows negligible temperature dependence below approximately 100~K, which justifies comparison with zero-temperature theoretical values.
Furthermore, the experimental EFG at Al(2) is closer to the Ce-based calculation than to that of the La analog, consistent with appreciable $4f$–ligand hybridization affecting the valence contribution to the EFG.
Similar estimations of Ce valence from $\nu_{\rm Q}$ variations, reflecting the itinerancy of the $4f$ electrons, have also been reported for the prototypical heavy-fermion compound CeCu$_2$Si$_2$ \cite{KobayashiTC:JPSJ82:2013}.


\subsection{Knight shifts and hyperfine coupling constants}\label{sec:KnightShift}

\begin{figure}[htb]
\includegraphics[width=8.5cm]{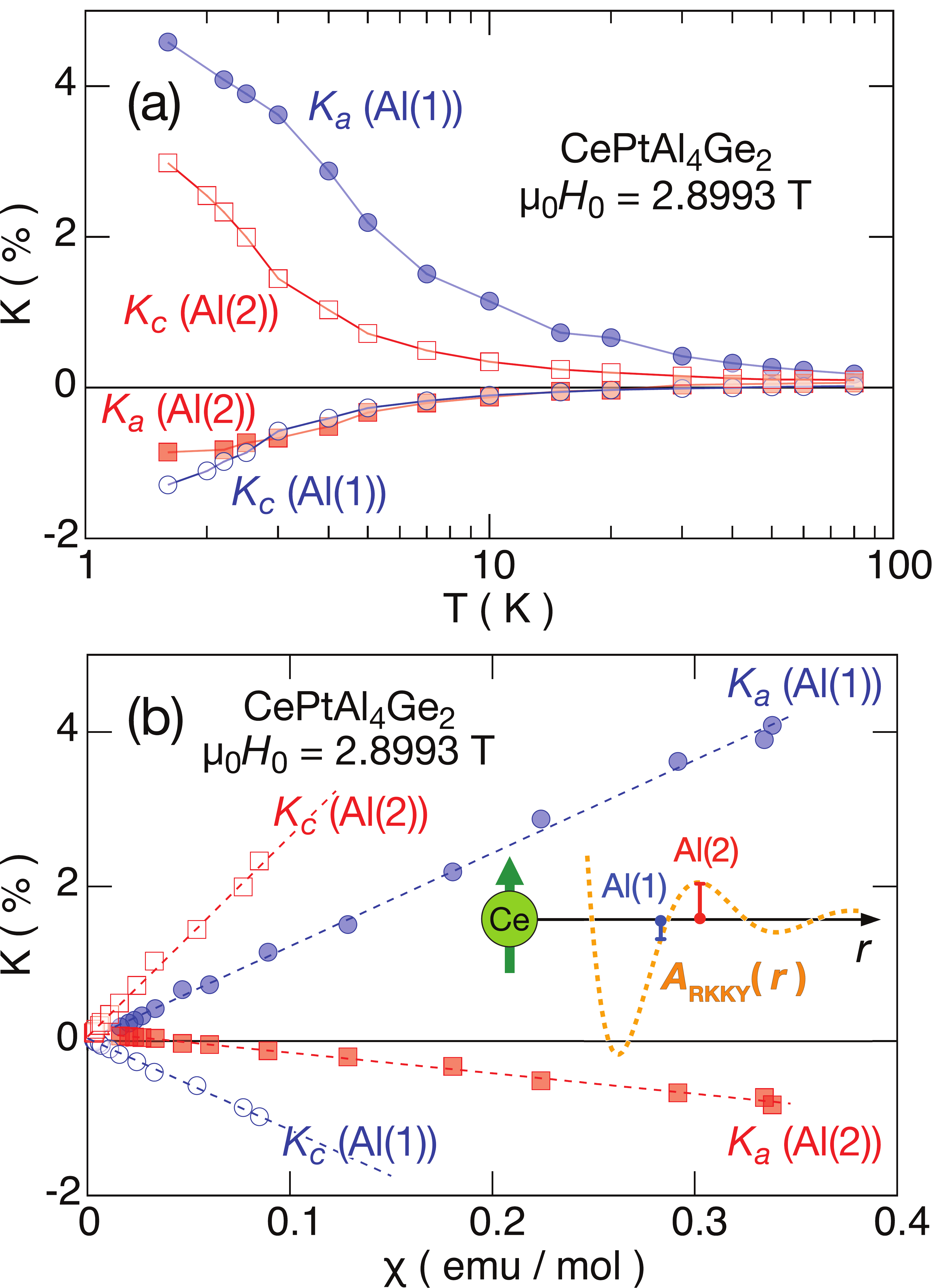}
\caption{\label{fig:KT_Kchi}(a) Temperature dependence of the Knight shifts $K_a$ and $K_c$ for the Al(1) and Al(2) sites in CePtAl$_4$Ge$_2$, measured at $\mu_0H_0 = 2.8993$ T.
(b) Knight shifts plotted against the bulk magnetic susceptibility $\chi$ in the paramagnetic state \cite{Shin2018Synthesis-and-c}, demonstrating linear $K$–$\chi$ relations.
The inset schematically illustrates the RKKY-type hyperfine coupling between the Ce and Al sites.}
\end{figure}

\begin{table*}[hbt]
\caption{\label{tab:NQR_hyperfine}
Summary of the experimental NQR frequency $\nu_{\rm Q}^{\rm exp}$, calculated values $\nu_{\rm Q}^{\rm calc}$ for LaPtAl$_4$Ge$_2$ and CePtAl$_4$Ge$_2$, transferred hyperfine coupling constants $A_a$ and $A_c$ (mT/$\mu_{\rm B}$), and the shortest Ce–Al bond length $d_{\rm Ce-Al}$ for the Al(1) and Al(2) sites in CePtAl$_4$Ge$_2$.
}
\begin{ruledtabular}
\begin{tabular}{lcccccc}
 \multirow{2}{*}{Site} & \multirow{2}{*}{$\nu_{\rm Q}^{\rm exp}$ (MHz)} & \multicolumn{2}{c}{$\nu_{\rm Q}^{\rm calc}$ (MHz)} & \multirow{2}{*}{$A_a$ (mT/$\mu_{\rm B}$)} & \multirow{2}{*}{$A_c$ (mT/$\mu_{\rm B}$)} & \multirow{2}{*}{$d_{\rm Ce-Al}$ (\AA)} \\
 {} & {} & {LaPtAl$_4$Ge$_2$} & {CePtAl$_4$Ge$_2$} & {} & {} & {} \\
\hline
Al(1) & $1.23 \pm 0.01$  & 1.261 & 1.278 & $-66 \pm 1$ & $67 \pm 1$ & 3.69 \\
Al(2) & $1.47 \pm 0.01$ & 1.368 & 1.461  & $-14.9 \pm 0.3$ & $145 \pm 3$ & 5.16 \\
\end{tabular}
\end{ruledtabular}
\end{table*}



Having established the site assignments of the two inequivalent Al sites, the temperature dependence of the Knight shifts $K_a$ and $K_c$ for Al(1) and Al(2) is obtained, as shown in Fig.~\ref{fig:KT_Kchi}(a).
For the Al(1) site, $K_a$ is positive and $K_c$ is negative, whereas the signs are reversed for the Al(2) site.
In all cases, the magnitude of the Knight shift increases monotonically upon cooling, following a Curie–Weiss-like behavior.
This indicates that the Knight shifts primarily reflect the $4f$ spin susceptibility of Ce, with distinct hyperfine couplings at each site.
Bulk magnetic susceptibility measurements \cite{Shin2018Synthesis-and-c} also reveal pronounced anisotropy, with $\chi_{ab} > \chi_c$, and show Curie–Weiss behavior down to $T_{\rm N}$ for $\chi_{ab}$, and above $\sim$275~K for $\chi_c$.
The effective moment is estimated as 2.5~$\mu_{\rm B}$ in the high temperature range, close to that of a free Ce$^{3+}$ ion.
The Weiss temperatures ($\theta_{\rm W}$) estimated from Curie–Weiss fits in the high temperature range are $-102$~K for $ H\parallel c$ and $38$~K for $ H\parallel a$.
The large magnitude of the Weiss temperature, relative to the N\'eel temperature, suggests the presence of strong magnetic frustration.

To derive the hyperfine interactions, standard $K$–$\chi$ plots are constructed using bulk magnetic susceptibility data and the Knight shift values for both crystallographic directions, as shown in Fig.~\ref{fig:KT_Kchi}(b).
For each Al site, linear relations are observed between the Knight shift and magnetic susceptibility, with the lines extrapolating to the origin.
The temperature-independent susceptibility is $\chi_0\approx0$ \cite{Shin2018Synthesis-and-c}, reflecting a near cancellation among Van Vleck paramagnetism, conduction-electron Pauli paramagnetism, a possible interband orbital contribution, Landau diamagnetism of the conduction electrons, and core-electron diamagnetism. The intercept $K_0$ in the $K$–$\chi$ plot collects the $T$-independent local contributions at the Al nucleus, namely a transferred Van Vleck term from the $4f$ ions together with conduction-electron Pauli and orbital terms at the ligand site.
The finding $K_0\approx0$ at $\chi=\chi_0\approx0$ in Fig.~\ref{fig:KT_Kchi}(b) indicates that these nonspin terms are negligible at the Al sites; consequently, the Al Knight shift can be regarded as arising almost entirely from the $4f$-moment contribution transferred via the hyperfine coupling.
The slopes of these $K$–$\chi$ plots yield the transferred hyperfine coupling constants $A_i$ along the $i$-axis ($i = a, c$), summarized in Table~\ref{tab:NQR_hyperfine}.
These values capture the anisotropic coupling between the Ce $4f$ electrons and the nuclear spins at the two distinct Al sites.

Despite the Al(1) site being located closer to the Ce atom than the Al(2) site [see Fig.~\ref{fig:CrystalStructure}(c) and Table~
\ref{tab:NQR_hyperfine}], the hyperfine coupling constants $|A_i|$ are notably larger at the Al(2) site.
This counterintuitive result suggests that the transferred hyperfine interaction is not simply determined by the Ce–Al distance ($r_i$), but is instead dominated by conduction-electron-mediated couplings, consistent with an RKKY-type mechanism, in which the magnitude of the transferred hyperfine field oscillates as a function of $r_i$.
In $f$-electron intermetallics, the hyperfine coupling along direction $i$ is generally expressed as $A_i=A_i^{\rm dip}+A_i^{\rm RKKY}$.
The dipolar term is
$A_i^{\rm dip}=\frac{\mu_0}{4\pi}\sum_j(3\cos^2\theta_{j,i}-1)/r_j^3$, where $\cos\theta_{j,i}=\hat{\bm{e}}_i\cdot\hat{\bm{r}}_j$, $r_j=|\bm{ r_j}|$, and $\hat{\bm{r}}_j=\bm{r_j}/r_j$.
The RKKY term represents the transferred hyperfine field mediated by conduction electrons; a minimal parametrization  is $A_i^{\rm RKKY}=C_i\sum_j w_{j,i}\,G(2k_{\rm F}r_j)$ with $G(x)=[x\cos x-\sin x]/x^4$.
Here $C_i\propto J_{cf,i}|u_{k_{\rm F}}(0)|^2$ sets the overall scale, including the usual Fermi-contact term, and $w_{j,i}$ encodes the geometry and the anisotropic 4$f$–conduction-electron hybridization along the Ce–Al paths.
Because $G(2k_{\rm F}r_j)$ is oscillatory and decays as $1/r_j^3$, $A_i^{\rm RKKY}$ changes sign and magnitude with $r_j$ and inherits directional anisotropy through $w_{j,i}$, as schematically illustrated in the inset of Fig.~\ref{fig:KT_Kchi}(b).
To evaluate the $A_i^{\rm dip}$, the expected dipolar fields were calculated by summing over Ce$^{3+}$ magnetic moments within a sphere of radius $\sim$100~\AA.
For the Al(1) site, the estimates yield $A_a^{\rm dip} \approx -28$~mT/$\mu_{\rm B}$ and $A_c^{\rm dip} \approx 64$~mT/$\mu_{\rm B}$, in good agreement with the experimental values.
This agreement suggests that the transferred $A^{\rm RKKY}$ is minimal at Al(1).
In contrast, the Al(2) site exhibits $A_a^{\rm dip} \approx -22$~mT/$\mu_{\rm B}$ and $A_c^{\rm dip} \approx 50$~mT/$\mu_{\rm B}$, which are much smaller than the measured values.
This substantial enhancement indicates a strong transferred hyperfine contribution $A^{\rm RKKY}$, reflecting pronounced $c$-axis anisotropy in the hybridization between Ce $4f$ electrons and conduction states at Al(2).

\subsection{Spin-lattice relaxation rate and spin dynamics in the paramagnetic state} \label{sec:invT1}

Spin-lattice relaxation rate ($1/T_1$) measurements of the $^{27}$Al nuclei were performed to probe the low-energy spin dynamics in CePtAl$_4$Ge$_2$ from a local microscopic perspective.
The nuclear magnetization recovery curves at all measured temperatures are well described by the fitting functions given in Sec.~\ref{sec:experimental}, confirming spatially uniform relaxation governed by magnetic fluctuations.
In general, the relaxation rate can be expressed as \cite{Moriya1963The-Effect-of-E}
\begin{equation}
\frac{1}{T_1} = 2 \left( \frac{\gamma_{\rm n}A_{\perp}}{\gamma_{\rm e}\hbar} \right)^2 k_{\rm B}T \sum_{\bm{q}} f^2(\bm{q}) \frac{{\rm Im}\chi_{\perp}(\bm{q},\omega_0)}{\omega_0},
\label{eq:invT1_new}
\end{equation}
where $\gamma_{\rm e}$ is the electronic gyromagnetic ratio, $A_{\perp}$ is the transverse component of the hyperfine coupling constant, $f(\bm{q})$ is the hyperfine form factor, ${\rm Im}\chi_{\perp}(\bm{q},\omega_0)$ is the imaginary part of the transverse dynamic spin susceptibility of the Ce $4f$ electrons, and $\omega_0$ is the Larmor angular frequency.
Since the ${\rm Im}\chi_{\perp}$ term reflects magnetic excitations perpendicular to $H_0$, the $1/T_1$ measurements selectively probe transverse low-energy spin fluctuations.

\begin{figure}[hbt]
\includegraphics[width=8.5cm]{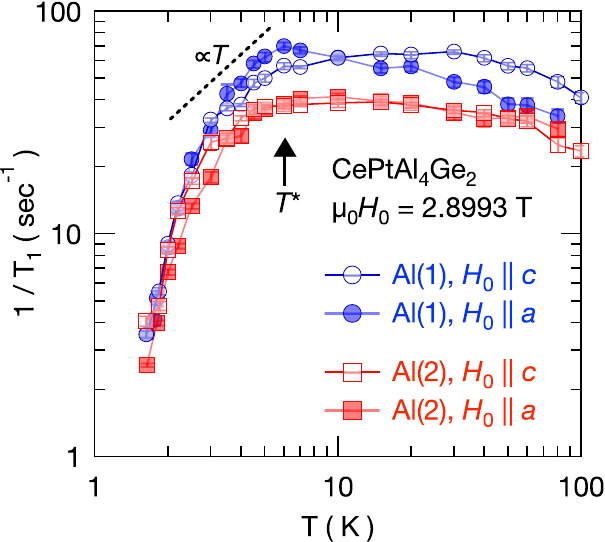}
\caption{\label{fig:invT1} Temperature dependence of the nuclear spin-lattice relaxation rate $1/T_1$ for the Al(1) and Al(2) sites in CePtAl$_4$Ge$_2$, measured under a magnetic field of $\mu_0H_0 = 2.8993$ T applied along the $a$ and $c$ axes.
The Kondo coherence temperature $T^\ast$, as reported by resistivity measurements~\cite{Shin2018Synthesis-and-c}, is indicated by the vertical arrow.}
\end{figure}

Figure~\ref{fig:invT1} shows the temperature dependence of $1/T_1$ for the Al(1) and Al(2) sites under a magnetic field of $\mu_0H_0 = 2.8993$~T applied along the $a$ and $c$ axes.
At this field strength, the AFM order is completely suppressed for both field orientations, and the system remains in the paramagnetic state over the entire temperature range studied [see Fig.~\ref{fig:NMR_AFM}(a)].
Above $T^{\ast} \approx 6$~K, $1/T_1$ exhibits only a weak temperature dependence, consistent with the behavior expected for localized Ce $4f$ moments.
This characteristic temperature $T^{\ast}$ coincides with the Kondo coherence temperature identified from resistivity measurements \cite{Shin2018Synthesis-and-c}.
Separately, the magnetic contribution to the resistivity, $\rho_{\rm m}$, obtained by subtracting the resistivity of LaPtAl$_4$Ge$_2$ from that of CePtAl$_4$Ge$_2$, exhibits another broad maximum around $\sim$120~K.
Bulk susceptibility measurements and neutron scattering results~\cite{Shin2020Magnetic-struct} further indicate that the crystalline electric field (CEF) ground state is the $|j_z = \pm 1/2\rangle$ Kramers doublet, with the first excited state $|j_z = \pm 3/2\rangle$ lying about 170~K higher in energy.
Therefore, in the temperature range between $T^{\ast}$ and $\sim$100~K, the spin dynamics are governed by localized moments in the CEF ground state undergoing Kondo interactions with conduction electrons.

In magnetic systems where localized moments are coupled via exchange interactions, Eq.~(\ref{eq:invT1_new}) can be simplified to the form proposed by Moriya \cite{Moriya_LocalMomentLimit}:
\begin{equation}
\left( \frac{1}{T_1} \right)_{\rm ex} = \frac{\sqrt{2} \pi \left( A_{\rm iso} / z' \right)^2 p_{\rm eff}^2}{3\hbar^2 \omega_{\rm ex}},
\label{eq:T1ex_new}
\end{equation}
where the exchange frequency $\omega_{\rm ex}$ is given by
\begin{equation}
\omega_{\rm ex}^2 = \frac{3k_{\rm B}^2 \theta_{\rm W}^2}{2z\,p_{\rm eff}^2 \hbar^2}.
\label{eq:omega_ex_new}
\end{equation}
Here, $A_{\rm iso}$ is the isotropic hyperfine coupling constant, $z'$ is the number of magnetic ions surrounding the nuclear site, $p_{\rm eff}$ is the effective magnetic moment, $z$ is the number of nearest-neighbor magnetic ions per Ce site, and $\theta_{\rm W}$ is the Weiss temperature.
By substituting $p_{\rm eff} \approx 1.8~\mu_{\rm B}$ estimated from the moment of the CEF ground state, and using $\theta_{\rm W} \approx 2.8$~K from Curie-Weiss fits for $H \parallel c$ in the low temperature range \cite{Shin2020Magnetic-struct}, the exchange frequency $\omega_{\rm ex}$ can be estimated.
For the Al(1) and Al(2) sites, we adopt $|A_{\rm iso}| = 67$~mT/$\mu_{\rm B}$ and $85$~mT/$\mu_{\rm B}$, with $z' = 3$ and $4$, respectively.
These values yield exchange-narrowed spin-lattice relaxation rates of $\left( 1/T_1 \right)_{\rm ex} \approx 43$~s$^{-1}$ for Al(1) and $\approx 39$~s$^{-1}$ for Al(2).
The close agreement between these estimates and the experimental $1/T_1$ values supports the applicability of the CEF ground-state configuration proposed in Ref.~\onlinecite{Shin2020Magnetic-struct} to the present system.

Upon the formation of Kondo lattice coherence, hybridization of localized $4f$ moments with conduction electrons would lead to a Korringa-like behavior, $1/T_1\propto T$.
However, in CePtAl$_4$Ge$_2$, $1/T_1$ decreases much more sharply than expected from this relation below $T^{\ast}\approx6$~K, with the rapid decrease starting at $T\sim3$–$4$~K, while $K(T)$ shows no corresponding anomaly, indicating a substantial reduction of low-energy spin fluctuations.
Fitting the data well below $T^{\ast}$ to an activated form, $1/T_1\propto \exp(-E_g/k_{\rm B}T)$, empirically yields an effective scale $E_g/k_{\rm B}\sim10$~K.
Given the limited $T$ window, no unique functional form is justified.
This behavior is consistent with a pseudogap- or spin-gap-like suppression at finite $\bm q$ (e.g., SDW-like partial gapping) rather than a uniform $\bm{q}=0$ density of states (DOS) gap.

\begin{figure}[bt]
\includegraphics[width=8.5cm]{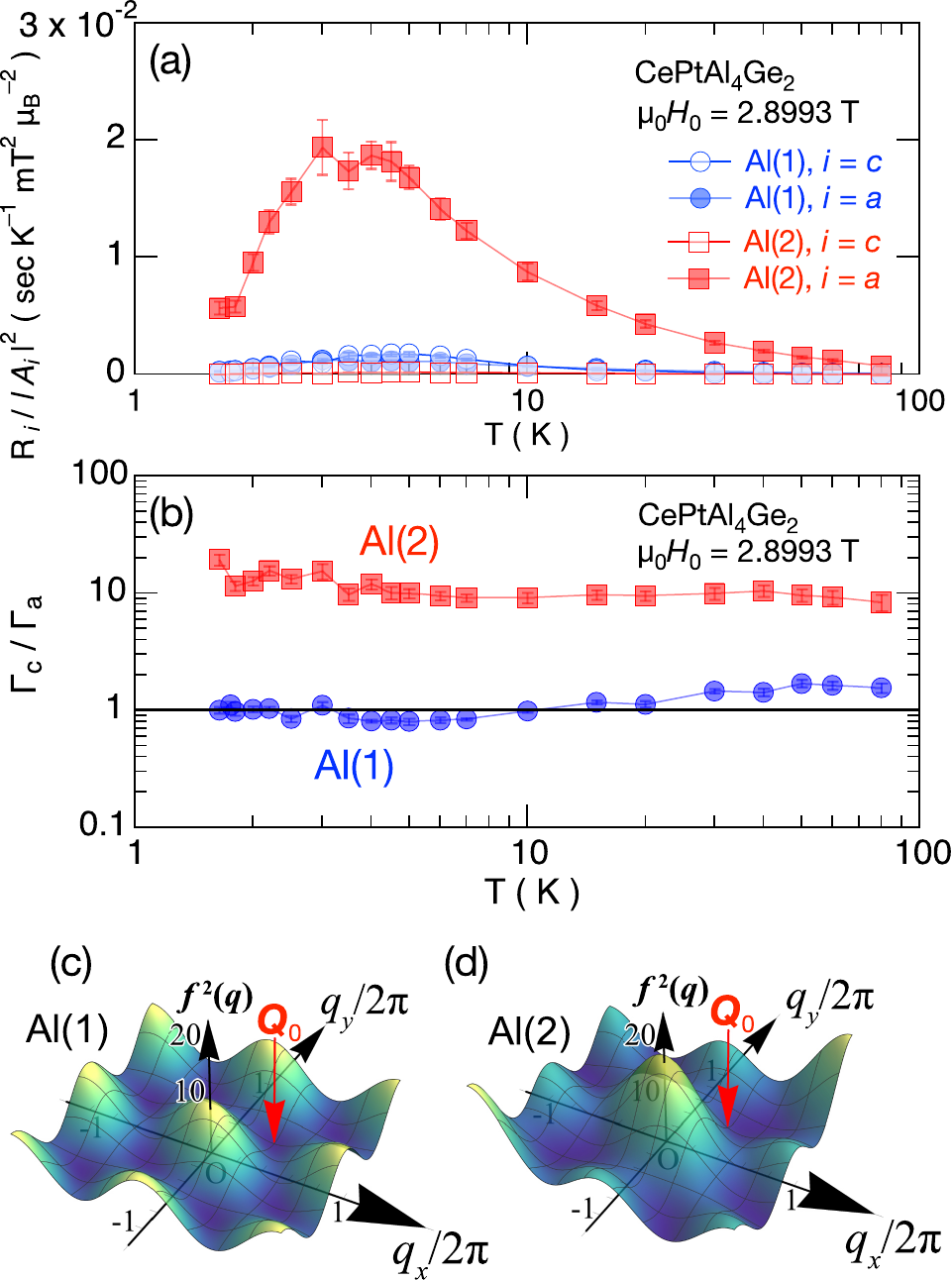}
\caption{\label{fig:Ri_Asq} (a) Temperature dependence of the fluctuation amplitude $R_i / |A_i|^2$ ($i=a, c$) at the Al(1) and Al(2) sites.
(b) Temperature dependence of the anisotropy ratio $\Gamma_c / \Gamma_a$ for spin fluctuations at each Al site. The quantity $\Gamma_i$ represents the characteristic energy scale of spin fluctuations along the $i$-axis; see main text for definition.
The hyperfine form factor $|f(\bm{q})|^2$ for (c) Al(1) and (d) Al(2) sites, respectively, are illustrated in the $q_x$–$q_y$ plane. Red arrows indicate the specific point at $\bm{Q}_0 = (\frac{1}{3}, \frac{1}{3})$. }
\end{figure}

To further investigate the nature of spin fluctuations, their anisotropy is analyzed by decomposing the fluctuation components along different crystallographic directions.
The fluctuation rate along the $i$-axis ($i = a$ or $c$) is defined as
\begin{equation}
R_i \equiv \left( \frac{\gamma_{\rm n} A_i}{\gamma_{\rm e} \hbar} \right)^2 k_{\rm B} \sum_{\bm{q}} f^2(\bm{q}) \frac{{\rm Im}\,\chi_i(\bm{q}, \omega_0)}{\omega_0}.
\end{equation}
From Eq.~(\ref{eq:invT1_new}), the quantities $R_i$ can be obtained experimentally as
$
R_a = \frac{1}{2}(T_1T)_{H_0 \parallel c}^{-1}$, $R_c = (T_1T)_{H_0 \parallel a}^{-1} - R_a$.
The fluctuation amplitude along the $i$-axis is then proportional to $R_i / |A_i|^2$.
Assuming a Lorentzian form for the dynamical spin susceptibility,
$\{{\rm Im}\,\chi_i(\bm{q}, \omega_0)\}/\omega_0 = \chi_i(\bm{q})/\Gamma_i(\bm{q})$,
where $\Gamma_i(\bm{q})$ is the characteristic energy scale of magnetic fluctuations, the $\bm{q}$-averaged fluctuation energy is defined as
$\Gamma_i \equiv \left[ \overline{\Gamma_i^2(\bm{q})} \right]^{1/2}$.
Within the strong correlation limit ($2\pi \chi_i(\bm{q}) \Gamma_i(\bm{q}) \sim 1$) \cite{MoriyaT:JPSJ64:1995}, and assuming $f^2(\bm{q}) = 1$, this reduces to
\begin{equation}
\Gamma_i = k_{\rm B}^{1/2} (\gamma_{\rm e} \hbar)^{-1} \left\{ \frac{\gamma_{\rm n} A_i}{(2\pi R_i)^{1/2}} \right\}.
\end{equation}
This approximation is generally valid when discussing spin fluctuation anisotropy at a given site \cite{Kambe_115T1anisotropy,SakaiH:PRB82:2010,BaekSH:PRL105:2010}, except in special cases where antiferromagnetic fluctuations are completely filtered out by symmetry at the nuclear site.

Figure~\ref{fig:Ri_Asq}(a) shows the temperature dependence of the fluctuation amplitude $R_i/|A_i|^2$ for the Al(1) and Al(2) sites.
For Al(2), the in-plane ($a$-axis) component is strongly enhanced relative to the $c$-axis component, whereas for Al(1) both components remain much smaller in magnitude.
Figure~\ref{fig:Ri_Asq}(b) presents the temperature dependence of the anisotropy ratio $\Gamma_c/\Gamma_a$, which reflects the ratio of the characteristic fluctuation energies along the two directions.
For Al(2), $\Gamma_c/\Gamma_a \approx 10$ with little temperature variation, indicating that $4f$ spin fluctuations in this temperature range are predominantly in-plane.
In contrast, the overall fluctuation amplitude for Al(1) is significantly smaller than the in-plane component for Al(2), despite both sites probing the same underlying $4f$ spin dynamics.
This pronounced disparity arises from the difference in hyperfine form factors $f(\bm{q})$ at the two sites, not meaning that the $4f$ spin fluctuations are different on the nuclear site-to-site.

Figures~\ref{fig:Ri_Asq}(c) and \ref{fig:Ri_Asq}(d) show the calculated $f^2(\bm{q})$ in the $q_x$–$q_y$ plane for Al(1) and Al(2).
As illustrated in Fig.~\ref{fig:CrystalStructure}(c), both sites lie above or below a Ce triangle, but Al(1) has two such triangles in the same Ce plane as nearest and next-nearest neighbors, whereas Al(2) has only one next-nearest Ce neighbor in the adjacent layer.
This distinct coordination yields different $\bm{q}$-dependent filtering of magnetic fluctuations: for example, $f^2(\bm{Q}_0)=0$ for Al(1) but $f^2(\bm{Q}_0)\neq0$ for Al(2) at $\bm{Q}_0=(\frac{1}{3}, \frac{1}{3})$, and similar contrasts occur at $\bm{Q}_1=(\frac{1}{2}, 0)$ and $\bm{Q}_2=(0, \frac{1}{2})$.
Neutron diffraction, however, finds an incommensurate ordering vector with in-plane component $\bm{Q}_{\rm AFM}=(1.39,0)$, for which $f^2(\bm{Q}_{\rm AFM})\neq0$ at both sites.
This implies that AFM fluctuations above $T_{\rm N}$ may initially emerge along high-symmetry $\bm{Q}_0$, $\bm{Q}_1$, or $\bm{Q}_2$ as in a localized-moment system, but in the itinerant $f$-electron case, the ordering vector is also shaped by the $\bm{k}$-dependent Fermi surface.
The combined influence of geometric frustration and itinerancy likely selects the incommensurate $\bm{Q}_{\rm AFM}$ over the nominal high-symmetry points.

In summary, $^{27}$Al NMR relaxation reveals strongly anisotropic spin fluctuations in CePtAl$_4$Ge$_2$.
Al(2) senses pronounced, nearly temperature-independent in-plane fluctuations, whereas Al(1) experiences much weaker ones due to hyperfine form factor filtering.
Thus, the dominant low-energy spin dynamics in the paramagnetic state are in-plane $4f$ AFM fluctuations whose momentum structure is strongly shaped by the triangular lattice network of Ce atoms.

\subsection{AFM order in CePtAl$_4$Ge$_2$}
\begin{figure}[bt]
\includegraphics[width=8.5cm]{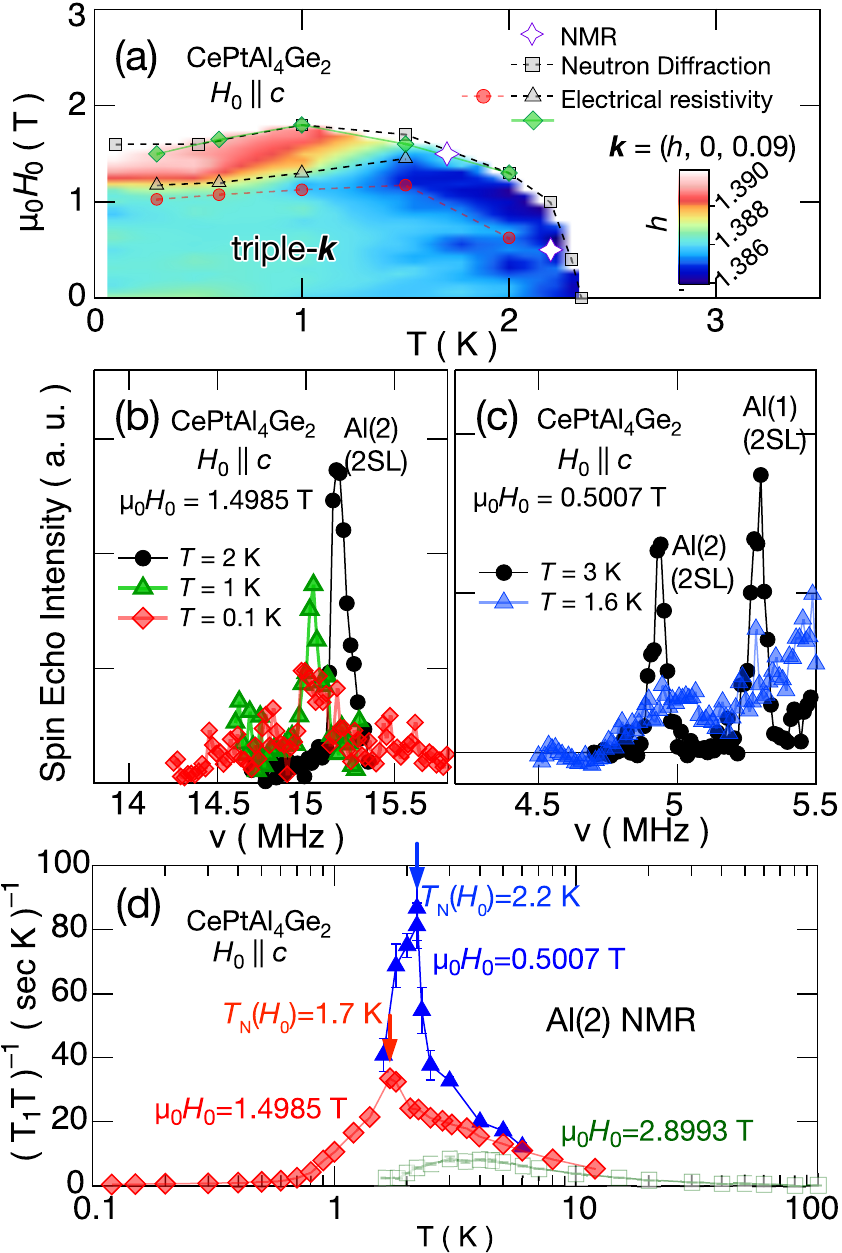}
\caption{\label{fig:NMR_AFM} (a) $H_0$–$T$ phase diagram of CePtAl$_4$Ge$_2$ for $H_0 \parallel c$, determined from neutron diffraction and electrical resistivity measurements~\cite{Shin_H-Tdiagram_preparation}.
The N\'eel temperatures $T_{\rm N}(H_0)$ determined from the peak in the NMR-$(T_1T)^{-1}$ [Fig.~\ref{fig:NMR_AFM}(d)] are also plotted.
The color contour represents the $h$-index of the incommensurate ordering vector $\bm{k} = (h, 0, 0.09)$.
(b) $^{27}$Al NMR spectra of the lower-frequency-side second satellite (2SL) transition of the Al(2) sites under $\mu_0H_0 = 1.4985$~T.
(c) $^{27}$Al NMR spectra of the 2SL transition of the Al(1) and Al(2) sites under $\mu_0H_0 = 0.5007$~T.
(d) Temperature dependence of $(T_1T)^{-1}$ at the Al(2) sites, measured under $\mu_0H_0 = 0.5007$, 1.4985, and 2.8993~T with $H_0 \parallel c$.
Vertical arrows indicate $T_{\rm N}(H_0)$ at each field.}
\end{figure}

As reported in Ref.~\onlinecite{Shin2020Magnetic-struct}, CePtAl$_4$Ge$_2$ undergoes an incommensurate longitudinal spin-density-wave (SDW) order — i.e., a spin-amplitude-modulated magnetic structure — with propagation vector $\bm{k}_{\rm AFM} \approx (1.39, 0, 0.09)$ at $T_{\rm N} = 2.3$~K in zero field.
Recent neutron diffraction measurements on single crystals further revealed that, in the low-field region, the magnetic structure adopts a triple-$\bm{k}$ configuration \cite{Shin2023Triple-sinusoid_ArXiv}.
In a triple-$\bm{k}$ state, three symmetry-equivalent modulation vectors $\bm{k}_1$, $\bm{k}_2$, and $\bm{k}_3$ contribute simultaneously, producing a noncollinear and noncoplanar spin texture that cannot be represented by a single propagation vector.
Neutron diffraction under $H_0 \parallel c$ suggests that a field-induced transition occurs, as shown in Fig.~\ref{fig:NMR_AFM}(a), accompanied by a slight change in the ordering wave vector and an abrupt suppression of the magnetic Bragg intensity \cite{Shin_H-Tdiagram_preparation}.
This may indicate a modification of the triple-$\bm{k}$ structure toward another amplitude-modulated configuration in the high-field AFM phase.
In both the zero-field triple-$\bm{k}$ AFM structure and the field-modified AFM state, which are forms of incommensurate longitudinal SDW order, the NMR spectra are expected to broaden significantly because the $^{27}$Al nuclei sense both the external field and site-dependent internal hyperfine fields arising from the spatial modulation of Ce moments.
As established in Sec.~\ref{sec:KnightShift}, the hyperfine coupling constant at the Al(2) site is larger than that at the Al(1) site, and as shown in Sec.\ref{sec:invT1}, the form factor does not suppress the internal field at relevant $\bm{q}$-vectors.
Therefore, the Al(2) site is particularly suitable for probing the AFM state by NMR.

Figure~\ref{fig:NMR_AFM}(b) shows the $^{27}$Al(2) NMR spectra at $\mu_0H_0 = 1.4985$~T ($\parallel\!\!c$), where the system enters the high-field AFM phase at low temperatures, as indicated in Fig.~\ref{fig:NMR_AFM}(a).
Below $T_{\rm N}(H_0)$, the spectra broaden markedly, and at $T = 0.1$~K they are substantially wider than in the paramagnetic state. The shift of the peak position below $T_{\rm N}(H_0)$ corresponds to a reduction of the Knight shift, reflecting the decrease in the macroscopic susceptibility upon the onset of AFM order. In such an SDW state, there are always positions where the internal field is canceled, and the spectrum broadens symmetrically around these zero internal field sites.
A similar broadening is observed at $\mu_0H_0 = 0.5007$~T [Fig.\ref{fig:NMR_AFM}(c)], although no reliable spectra could be obtained below 1.6~K.
This loss of signal likely originates from the combined effects of reduced NMR signal intensity at low fields and pronounced line broadening associated with the triple-$\bm{k}$ structure.
The inability to detect signals below 1.6~K in low fields suggests that the sensitivity of the present $^3$He–$^4$He dilution-refrigerator NMR setup was insufficient for such conditions.
As demonstrated in the detailed NMR study of the polar magnet GaV$_4$Se$_8$ hosting a N\'eel-type skyrmion lattice \cite{Takeda2024Magnetic-struct}, future investigations of the triple-$\bm{k}$ structure will require larger single crystals and improved probe sensitivity.

The $(T_1T)^{-1}$ data for the Al(2) site at various fields with $H_0\!\parallel\!c$ are summarized in Fig.~\ref{fig:NMR_AFM}(d) and, by Eq.~(\ref{fig:invT1}), probe in-plane spin fluctuations.
At $\mu_0H_0\!\approx\!2.9$~T, $(T_1T)^{-1}$ exhibits a broad maximum at 3–4~K below $T^{\ast}\!\sim\!6$~K, indicative of a pseudogap-like suppression associated with the onset of hybridization of $4f$ electrons with conduction electrons.
At $\mu_0H_0\!\approx\!1.5$~T, $(T_1T)^{-1}$ increases monotonically on cooling from $\sim\!10$~K toward $T_{\rm N}(H_0)$, where a sharp maximum appears due to critical slowing down at $\bm q\!\approx\!\bm Q$; below $T_{\rm N}$, $(T_1T)^{-1}$ drops rapidly, consistent with the opening of a gap either in the low-energy spin-excitation spectrum or as an SDW gap on the Fermi surface.
The resulting transition temperatures agree with the $H_0$–$T$ phase diagram in Fig.~\ref{fig:NMR_AFM}(a).
Further reducing the field to $\mu_0H_0\!\approx\!0.5$~T produces an even stronger enhancement of $(T_1T)^{-1}$ on cooling, signaling a marked growth of in-plane fluctuations as the system approaches the zero-field triple-$\bm k$ AFM state.
These trends indicate that in-plane spin fluctuations strengthen toward low fields, consistent with their key role in stabilizing the triple-$\bm k$ AFM order.

\section{Conclusion}
A comprehensive site-selective $^{27}$Al NMR investigation has been carried out on the triangular Kondo lattice compound CePtAl$_4$Ge$_2$, which exhibits incommensurate antiferromagnetic order below $T_{\rm N} = 2.3$~K.
The two inequivalent Al sites were unambiguously assigned through comparisons of experimental quadrupolar parameters with the calculated EFG from the electronic structure calculation, enabling the determination of site-resolved hyperfine coupling constants.
These results show that the magnetic couplings are dominated by transferred RKKY-type interactions rather than simple dipolar fields.
In the paramagnetic state, $1/T_1$ measurements revealed pronounced anisotropic spin fluctuations with a predominantly in-plane character.
Below the Kondo coherence temperature $T^{\ast} \approx 6$~K, $1/T_1$ decreases much more steeply than expected from Korringa behavior, suggesting the opening of a spin gap of order $\sim 10$~K as the Ce $4f$ moments evolve into a coherent heavy Fermi liquid.
Comparison of fluctuation amplitudes at the two Al sites demonstrated that the apparent site dependence originates from the distinct hyperfine form factors $f(\bm{q})$.
The incommensurate ordering vector $\bm{Q}_{\rm AFM}$ determined by neutron diffraction yields finite $f^2(\bm{Q}_{\rm AFM})$ at both sites, indicating that long-range order develops at a geometrically compromised wave vector selected from underlying frustrated correlations.

In the ordered state, NMR spectra broaden substantially and $(T_1T)^{-1}$ drops sharply below $T_{\rm N}(H_0)$, consistent with the onset of long-range magnetic order.
The magnetic-field dependence of $(T_1T)^{-1}$ reveals that in-plane spin fluctuations are strongly enhanced near $T_{\rm N}$, particularly at low fields.

These findings establish CePtAl$_4$Ge$_2$ as a clean platform for studying frustrated Kondo magnetism in a triangular lattice network, and demonstrate the effectiveness of site-selective NMR in resolving momentum-dependent spin dynamics and complex spin textures in correlated $f$-electron systems.
The predominance of in-plane spin fluctuations found here bears a notable resemblance to that observed in the triangular Kondo lattice compound YbV$_6$Sn$_6$, suggesting a common underlying mechanism arising from the synergy of Kondo physics and geometrical frustration in layered $f$-electron materials \cite{Park2025Investigation-o}.
Taken together, our results on CePtAl$_4$Ge$_2$ provide microscopic insight into how geometric frustration and Kondo coherence intertwine to shape unconventional magnetism in $f$-electron systems.

\begin{acknowledgments}
We thank T. Kitazawa and Y. Haga for their assistance with single-crystal orientation and for fruitful discussions.
This work was supported by the Japan Society for the Promotion of Science (JSPS) through KAKENHI (Grants No. JP23K25829, No. JP24KK0062, and No. JP24K00587).
Work at the Japan Atomic Energy Agency (JAEA) was partially supported by the JAEA REIMEI Research Program.
S. S. acknowledges support from the Swiss National Science Foundation SNSF Project No. 188706.
Work at Sungkyunkwan University was supported by the National Research Foundation (NRF) of Korea through a grant funded by the Korean government (Grants No. RS-2023-00220471 and RS-2021-NR059409).
\end{acknowledgments}


\begin{thebibliography}{32}%
\makeatletter
\providecommand \@ifxundefined [1]{%
 \@ifx{#1\undefined}
}%
\providecommand \@ifnum [1]{%
 \ifnum #1\expandafter \@firstoftwo
 \else \expandafter \@secondoftwo
 \fi
}%
\providecommand \@ifx [1]{%
 \ifx #1\expandafter \@firstoftwo
 \else \expandafter \@secondoftwo
 \fi
}%
\providecommand \natexlab [1]{#1}%
\providecommand \enquote  [1]{``#1''}%
\providecommand \bibnamefont  [1]{#1}%
\providecommand \bibfnamefont [1]{#1}%
\providecommand \citenamefont [1]{#1}%
\providecommand \href@noop [0]{\@secondoftwo}%
\providecommand \href [0]{\begingroup \@sanitize@url \@href}%
\providecommand \@href[1]{\@@startlink{#1}\@@href}%
\providecommand \@@href[1]{\endgroup#1\@@endlink}%
\providecommand \@sanitize@url [0]{\catcode `\\12\catcode `\$12\catcode `\&12\catcode `\#12\catcode `\^12\catcode `\_12\catcode `\%12\relax}%
\providecommand \@@startlink[1]{}%
\providecommand \@@endlink[0]{}%
\providecommand \url  [0]{\begingroup\@sanitize@url \@url }%
\providecommand \@url [1]{\endgroup\@href {#1}{\urlprefix }}%
\providecommand \urlprefix  [0]{URL }%
\providecommand \Eprint [0]{\href }%
\providecommand \doibase [0]{https://doi.org/}%
\providecommand \selectlanguage [0]{\@gobble}%
\providecommand \bibinfo  [0]{\@secondoftwo}%
\providecommand \bibfield  [0]{\@secondoftwo}%
\providecommand \translation [1]{[#1]}%
\providecommand \BibitemOpen [0]{}%
\providecommand \bibitemStop [0]{}%
\providecommand \bibitemNoStop [0]{.\EOS\space}%
\providecommand \EOS [0]{\spacefactor3000\relax}%
\providecommand \BibitemShut  [1]{\csname bibitem#1\endcsname}%
\let\auto@bib@innerbib\@empty
\bibitem [{\citenamefont {Si}(2006)}]{Si2006Global-magnetic}%
  \BibitemOpen
  \bibfield  {author} {\bibinfo {author} {\bibfnamefont {Q.}~\bibnamefont {Si}},\ }\bibfield  {title} {\bibinfo {title} {Global magnetic phase diagram and local quantum criticality in heavy fermion metals},\ }\href {https://doi.org/https://doi.org/10.1016/j.physb.2006.01.156} {\bibfield  {journal} {\bibinfo  {journal} {Physica B: Condensed Matter}\ }\textbf {\bibinfo {volume} {378-380}},\ \bibinfo {pages} {23} (\bibinfo {year} {2006})}\BibitemShut {NoStop}%
\bibitem [{\citenamefont {Vojta}(2008)}]{Vojta2008From-itinerant-}%
  \BibitemOpen
  \bibfield  {author} {\bibinfo {author} {\bibfnamefont {M.}~\bibnamefont {Vojta}},\ }\bibfield  {title} {\bibinfo {title} {From itinerant to local-moment antiferromagnetism in kondo lattices: Adiabatic continuity versus quantum phase transitions},\ }\href {https://doi.org/10.1103/PhysRevB.78.125109} {\bibfield  {journal} {\bibinfo  {journal} {Phys. Rev. B}\ }\textbf {\bibinfo {volume} {78}},\ \bibinfo {pages} {125109} (\bibinfo {year} {2008})}\BibitemShut {NoStop}%
\bibitem [{\citenamefont {Coleman}\ and\ \citenamefont {Nevidomskyy}(2010)}]{Coleman2010Frustration-and}%
  \BibitemOpen
  \bibfield  {author} {\bibinfo {author} {\bibfnamefont {P.}~\bibnamefont {Coleman}}\ and\ \bibinfo {author} {\bibfnamefont {A.~H.}\ \bibnamefont {Nevidomskyy}},\ }\bibfield  {title} {\bibinfo {title} {Frustration and the kondo effect in heavy fermion materials},\ }\href {https://doi.org/10.1007/s10909-010-0213-4} {\bibfield  {journal} {\bibinfo  {journal} {Journal of Low Temperature Physics}\ }\textbf {\bibinfo {volume} {161}},\ \bibinfo {pages} {182} (\bibinfo {year} {2010})}\BibitemShut {NoStop}%
\bibitem [{\citenamefont {Batista}\ \emph {et~al.}(2016)\citenamefont {Batista}, \citenamefont {Lin}, \citenamefont {Hayami},\ and\ \citenamefont {Kamiya}}]{Batista2016Frustration-and}%
  \BibitemOpen
  \bibfield  {author} {\bibinfo {author} {\bibfnamefont {C.~D.}\ \bibnamefont {Batista}}, \bibinfo {author} {\bibfnamefont {S.-Z.}\ \bibnamefont {Lin}}, \bibinfo {author} {\bibfnamefont {S.}~\bibnamefont {Hayami}},\ and\ \bibinfo {author} {\bibfnamefont {Y.}~\bibnamefont {Kamiya}},\ }\bibfield  {title} {\bibinfo {title} {Frustration and chiral orderings in correlated electron systems},\ }\href {https://doi.org/10.1088/0034-4885/79/8/084504} {\bibfield  {journal} {\bibinfo  {journal} {Reports on Progress in Physics}\ }\textbf {\bibinfo {volume} {79}},\ \bibinfo {pages} {084504} (\bibinfo {year} {2016})}\BibitemShut {NoStop}%
\bibitem [{\citenamefont {D{\"o}nni}\ \emph {et~al.}(1996)\citenamefont {D{\"o}nni}, \citenamefont {Ehlers}, \citenamefont {Maletta}, \citenamefont {Fischer}, \citenamefont {Kitazawa},\ and\ \citenamefont {Zolliker}}]{Donni1996Geometrically-f}%
  \BibitemOpen
  \bibfield  {author} {\bibinfo {author} {\bibfnamefont {A.}~\bibnamefont {D{\"o}nni}}, \bibinfo {author} {\bibfnamefont {G.}~\bibnamefont {Ehlers}}, \bibinfo {author} {\bibfnamefont {H.}~\bibnamefont {Maletta}}, \bibinfo {author} {\bibfnamefont {P.}~\bibnamefont {Fischer}}, \bibinfo {author} {\bibfnamefont {H.}~\bibnamefont {Kitazawa}},\ and\ \bibinfo {author} {\bibfnamefont {M.}~\bibnamefont {Zolliker}},\ }\bibfield  {title} {\bibinfo {title} {Geometrically frustrated magnetic structures of the heavy-fermion compound {CePdAl} studied by powder neutron diffraction},\ }\href {https://doi.org/10.1088/0953-8984/8/50/043} {\bibfield  {journal} {\bibinfo  {journal} {Journal of Physics: Condensed Matter}\ }\textbf {\bibinfo {volume} {8}},\ \bibinfo {pages} {11213} (\bibinfo {year} {1996})}\BibitemShut {NoStop}%
\bibitem [{\citenamefont {Oyamada}\ \emph {et~al.}(2008)\citenamefont {Oyamada}, \citenamefont {Maegawa}, \citenamefont {Nishiyama}, \citenamefont {Kitazawa},\ and\ \citenamefont {Isikawa}}]{Oyamada2008Ordering-mechan}%
  \BibitemOpen
  \bibfield  {author} {\bibinfo {author} {\bibfnamefont {A.}~\bibnamefont {Oyamada}}, \bibinfo {author} {\bibfnamefont {S.}~\bibnamefont {Maegawa}}, \bibinfo {author} {\bibfnamefont {M.}~\bibnamefont {Nishiyama}}, \bibinfo {author} {\bibfnamefont {H.}~\bibnamefont {Kitazawa}},\ and\ \bibinfo {author} {\bibfnamefont {Y.}~\bibnamefont {Isikawa}},\ }\bibfield  {title} {\bibinfo {title} {Ordering mechanism and spin fluctuations in a geometrically frustrated heavy-fermion antiferromagnet on the kagome-like lattice {CePdAl}: A $^{27}${Al} {NMR} study},\ }\href {https://doi.org/10.1103/PhysRevB.77.064432} {\bibfield  {journal} {\bibinfo  {journal} {Phys. Rev. B}\ }\textbf {\bibinfo {volume} {77}},\ \bibinfo {pages} {064432} (\bibinfo {year} {2008})}\BibitemShut {NoStop}%
\bibitem [{\citenamefont {Zhao}\ \emph {et~al.}(2016)\citenamefont {Zhao}, \citenamefont {Zhang}, \citenamefont {Hu}, \citenamefont {Isikawa}, \citenamefont {Luo}, \citenamefont {Steglich},\ and\ \citenamefont {Sun}}]{Zhao2016Temperature-fie}%
  \BibitemOpen
  \bibfield  {author} {\bibinfo {author} {\bibfnamefont {H.}~\bibnamefont {Zhao}}, \bibinfo {author} {\bibfnamefont {J.}~\bibnamefont {Zhang}}, \bibinfo {author} {\bibfnamefont {S.}~\bibnamefont {Hu}}, \bibinfo {author} {\bibfnamefont {Y.}~\bibnamefont {Isikawa}}, \bibinfo {author} {\bibfnamefont {J.}~\bibnamefont {Luo}}, \bibinfo {author} {\bibfnamefont {F.}~\bibnamefont {Steglich}},\ and\ \bibinfo {author} {\bibfnamefont {P.}~\bibnamefont {Sun}},\ }\bibfield  {title} {\bibinfo {title} {Temperature-field phase diagram of geometrically frustrated {CePdAl}},\ }\href {https://doi.org/10.1103/PhysRevB.94.235131} {\bibfield  {journal} {\bibinfo  {journal} {Phys. Rev. B}\ }\textbf {\bibinfo {volume} {94}},\ \bibinfo {pages} {235131} (\bibinfo {year} {2016})}\BibitemShut {NoStop}%
\bibitem [{\citenamefont {Lucas}\ \emph {et~al.}(2017)\citenamefont {Lucas}, \citenamefont {Grube}, \citenamefont {Huang}, \citenamefont {Sakai}, \citenamefont {Wunderlich}, \citenamefont {Green}, \citenamefont {Wosnitza}, \citenamefont {Fritsch}, \citenamefont {Gegenwart}, \citenamefont {Stockert},\ and\ \citenamefont {v.~L\"ohneysen}}]{Lucas2017Entropy-Evoluti}%
  \BibitemOpen
  \bibfield  {author} {\bibinfo {author} {\bibfnamefont {S.}~\bibnamefont {Lucas}}, \bibinfo {author} {\bibfnamefont {K.}~\bibnamefont {Grube}}, \bibinfo {author} {\bibfnamefont {C.-L.}\ \bibnamefont {Huang}}, \bibinfo {author} {\bibfnamefont {A.}~\bibnamefont {Sakai}}, \bibinfo {author} {\bibfnamefont {S.}~\bibnamefont {Wunderlich}}, \bibinfo {author} {\bibfnamefont {E.~L.}\ \bibnamefont {Green}}, \bibinfo {author} {\bibfnamefont {J.}~\bibnamefont {Wosnitza}}, \bibinfo {author} {\bibfnamefont {V.}~\bibnamefont {Fritsch}}, \bibinfo {author} {\bibfnamefont {P.}~\bibnamefont {Gegenwart}}, \bibinfo {author} {\bibfnamefont {O.}~\bibnamefont {Stockert}},\ and\ \bibinfo {author} {\bibfnamefont {H.}~\bibnamefont {v.~L\"ohneysen}},\ }\bibfield  {title} {\bibinfo {title} {Entropy evolution in the magnetic phases of partially frustrated {CePdAl}},\ }\href {https://doi.org/10.1103/PhysRevLett.118.107204} {\bibfield  {journal} {\bibinfo  {journal} {Phys. Rev. Lett.}\ }\textbf {\bibinfo {volume} {118}},\ \bibinfo {pages}
  {107204} (\bibinfo {year} {2017})}\BibitemShut {NoStop}%
\bibitem [{\citenamefont {Mentink}\ \emph {et~al.}(1994)\citenamefont {Mentink}, \citenamefont {Drost}, \citenamefont {Nieuwenhuys}, \citenamefont {Frikkee}, \citenamefont {Menovsky},\ and\ \citenamefont {Mydosh}}]{Mentink1994Magnetic-Orderi}%
  \BibitemOpen
  \bibfield  {author} {\bibinfo {author} {\bibfnamefont {S.~A.~M.}\ \bibnamefont {Mentink}}, \bibinfo {author} {\bibfnamefont {A.}~\bibnamefont {Drost}}, \bibinfo {author} {\bibfnamefont {G.~J.}\ \bibnamefont {Nieuwenhuys}}, \bibinfo {author} {\bibfnamefont {E.}~\bibnamefont {Frikkee}}, \bibinfo {author} {\bibfnamefont {A.~A.}\ \bibnamefont {Menovsky}},\ and\ \bibinfo {author} {\bibfnamefont {J.~A.}\ \bibnamefont {Mydosh}},\ }\bibfield  {title} {\bibinfo {title} {Magnetic ordering and frustration in hexagonal {UNi}$_{4}${B}},\ }\href {https://doi.org/10.1103/PhysRevLett.73.1031} {\bibfield  {journal} {\bibinfo  {journal} {Phys. Rev. Lett.}\ }\textbf {\bibinfo {volume} {73}},\ \bibinfo {pages} {1031} (\bibinfo {year} {1994})}\BibitemShut {NoStop}%
\bibitem [{\citenamefont {Mentink}\ \emph {et~al.}(1995)\citenamefont {Mentink}, \citenamefont {Nieuwenhuys}, \citenamefont {Nakotte}, \citenamefont {Menovsky}, \citenamefont {Drost}, \citenamefont {Frikkee},\ and\ \citenamefont {Mydosh}}]{Mentink1995Magnetization-a}%
  \BibitemOpen
  \bibfield  {author} {\bibinfo {author} {\bibfnamefont {S.~A.~M.}\ \bibnamefont {Mentink}}, \bibinfo {author} {\bibfnamefont {G.~J.}\ \bibnamefont {Nieuwenhuys}}, \bibinfo {author} {\bibfnamefont {H.}~\bibnamefont {Nakotte}}, \bibinfo {author} {\bibfnamefont {A.~A.}\ \bibnamefont {Menovsky}}, \bibinfo {author} {\bibfnamefont {A.}~\bibnamefont {Drost}}, \bibinfo {author} {\bibfnamefont {E.}~\bibnamefont {Frikkee}},\ and\ \bibinfo {author} {\bibfnamefont {J.~A.}\ \bibnamefont {Mydosh}},\ }\bibfield  {title} {\bibinfo {title} {Magnetization and resistivity of {UNi}$_{4}${B} in high magnetic fields},\ }\href {https://doi.org/10.1103/PhysRevB.51.11567} {\bibfield  {journal} {\bibinfo  {journal} {Phys. Rev. B}\ }\textbf {\bibinfo {volume} {51}},\ \bibinfo {pages} {11567} (\bibinfo {year} {1995})}\BibitemShut {NoStop}%
\bibitem [{\citenamefont {Movshovich}\ \emph {et~al.}(1999)\citenamefont {Movshovich}, \citenamefont {Jaime}, \citenamefont {Mentink}, \citenamefont {Menovsky},\ and\ \citenamefont {Mydosh}}]{Movshovich1999Second-Low-Temp}%
  \BibitemOpen
  \bibfield  {author} {\bibinfo {author} {\bibfnamefont {R.}~\bibnamefont {Movshovich}}, \bibinfo {author} {\bibfnamefont {M.}~\bibnamefont {Jaime}}, \bibinfo {author} {\bibfnamefont {S.}~\bibnamefont {Mentink}}, \bibinfo {author} {\bibfnamefont {A.~A.}\ \bibnamefont {Menovsky}},\ and\ \bibinfo {author} {\bibfnamefont {J.~A.}\ \bibnamefont {Mydosh}},\ }\bibfield  {title} {\bibinfo {title} {Second low-temperature phase transition in frustrated {UNi}$_{4}${B}},\ }\href {https://doi.org/10.1103/PhysRevLett.83.2065} {\bibfield  {journal} {\bibinfo  {journal} {Phys. Rev. Lett.}\ }\textbf {\bibinfo {volume} {83}},\ \bibinfo {pages} {2065} (\bibinfo {year} {1999})}\BibitemShut {NoStop}%
\bibitem [{\citenamefont {Shin}\ \emph {et~al.}(2018)\citenamefont {Shin}, \citenamefont {Rosa}, \citenamefont {Ronning}, \citenamefont {Thompson}, \citenamefont {Scott}, \citenamefont {Lee}, \citenamefont {Jang}, \citenamefont {Jung}, \citenamefont {Yun}, \citenamefont {Lee}, \citenamefont {Bauer},\ and\ \citenamefont {Park}}]{Shin2018Synthesis-and-c}%
  \BibitemOpen
  \bibfield  {author} {\bibinfo {author} {\bibfnamefont {S.}~\bibnamefont {Shin}}, \bibinfo {author} {\bibfnamefont {P.~F.}\ \bibnamefont {Rosa}}, \bibinfo {author} {\bibfnamefont {F.}~\bibnamefont {Ronning}}, \bibinfo {author} {\bibfnamefont {J.~D.}\ \bibnamefont {Thompson}}, \bibinfo {author} {\bibfnamefont {B.~L.}\ \bibnamefont {Scott}}, \bibinfo {author} {\bibfnamefont {S.}~\bibnamefont {Lee}}, \bibinfo {author} {\bibfnamefont {H.}~\bibnamefont {Jang}}, \bibinfo {author} {\bibfnamefont {S.-G.}\ \bibnamefont {Jung}}, \bibinfo {author} {\bibfnamefont {E.}~\bibnamefont {Yun}}, \bibinfo {author} {\bibfnamefont {H.}~\bibnamefont {Lee}}, \bibinfo {author} {\bibfnamefont {E.~D.}\ \bibnamefont {Bauer}},\ and\ \bibinfo {author} {\bibfnamefont {T.}~\bibnamefont {Park}},\ }\bibfield  {title} {\bibinfo {title} {Synthesis and characterization of the heavy-fermion compound {CePtAl}$_4${Ge}$_2$},\ }\href {https://doi.org/https://doi.org/10.1016/j.jallcom.2017.12.180} {\bibfield  {journal} {\bibinfo  {journal} {Journal
  of Alloys and Compounds}\ }\textbf {\bibinfo {volume} {738}},\ \bibinfo {pages} {550} (\bibinfo {year} {2018})}\BibitemShut {NoStop}%
\bibitem [{\citenamefont {Shin}\ \emph {et~al.}(2020)\citenamefont {Shin}, \citenamefont {Pomjakushin}, \citenamefont {Keller}, \citenamefont {Rosa}, \citenamefont {Stuhr}, \citenamefont {Niedermayer}, \citenamefont {Sibille}, \citenamefont {Toth}, \citenamefont {Kim}, \citenamefont {Jang}, \citenamefont {Son}, \citenamefont {Lee}, \citenamefont {Shang}, \citenamefont {Medarde}, \citenamefont {Bauer}, \citenamefont {Kenzelmann},\ and\ \citenamefont {Park}}]{Shin2020Magnetic-struct}%
  \BibitemOpen
  \bibfield  {author} {\bibinfo {author} {\bibfnamefont {S.}~\bibnamefont {Shin}}, \bibinfo {author} {\bibfnamefont {V.}~\bibnamefont {Pomjakushin}}, \bibinfo {author} {\bibfnamefont {L.}~\bibnamefont {Keller}}, \bibinfo {author} {\bibfnamefont {P.~F.~S.}\ \bibnamefont {Rosa}}, \bibinfo {author} {\bibfnamefont {U.}~\bibnamefont {Stuhr}}, \bibinfo {author} {\bibfnamefont {C.}~\bibnamefont {Niedermayer}}, \bibinfo {author} {\bibfnamefont {R.}~\bibnamefont {Sibille}}, \bibinfo {author} {\bibfnamefont {S.}~\bibnamefont {Toth}}, \bibinfo {author} {\bibfnamefont {J.}~\bibnamefont {Kim}}, \bibinfo {author} {\bibfnamefont {H.}~\bibnamefont {Jang}}, \bibinfo {author} {\bibfnamefont {S.-K.}\ \bibnamefont {Son}}, \bibinfo {author} {\bibfnamefont {H.-O.}\ \bibnamefont {Lee}}, \bibinfo {author} {\bibfnamefont {T.}~\bibnamefont {Shang}}, \bibinfo {author} {\bibfnamefont {M.}~\bibnamefont {Medarde}}, \bibinfo {author} {\bibfnamefont {E.~D.}\ \bibnamefont {Bauer}}, \bibinfo {author} {\bibfnamefont {M.}~\bibnamefont
  {Kenzelmann}},\ and\ \bibinfo {author} {\bibfnamefont {T.}~\bibnamefont {Park}},\ }\bibfield  {title} {\bibinfo {title} {Magnetic structure and crystalline electric field effects in the triangular antiferromagnet {CePtAl}$_{4}${Ge}$_{2}$},\ }\href {https://doi.org/10.1103/PhysRevB.101.224421} {\bibfield  {journal} {\bibinfo  {journal} {Phys. Rev. B}\ }\textbf {\bibinfo {volume} {101}},\ \bibinfo {pages} {224421} (\bibinfo {year} {2020})}\BibitemShut {NoStop}%
\bibitem [{\citenamefont {Shin}\ \emph {et~al.}(2023)\citenamefont {Shin}, \citenamefont {Park}, \citenamefont {Sibille}, \citenamefont {Jang}, \citenamefont {Park}, \citenamefont {Kim}, \citenamefont {Shang}, \citenamefont {Medarde}, \citenamefont {Bauer}, \citenamefont {Zaharko}, \citenamefont {Kenzelmann},\ and\ \citenamefont {Park}}]{Shin2023Triple-sinusoid_ArXiv}%
  \BibitemOpen
  \bibfield  {author} {\bibinfo {author} {\bibfnamefont {S.}~\bibnamefont {Shin}}, \bibinfo {author} {\bibfnamefont {J.-H.}\ \bibnamefont {Park}}, \bibinfo {author} {\bibfnamefont {R.}~\bibnamefont {Sibille}}, \bibinfo {author} {\bibfnamefont {H.}~\bibnamefont {Jang}}, \bibinfo {author} {\bibfnamefont {T.~B.}\ \bibnamefont {Park}}, \bibinfo {author} {\bibfnamefont {S.}~\bibnamefont {Kim}}, \bibinfo {author} {\bibfnamefont {T.}~\bibnamefont {Shang}}, \bibinfo {author} {\bibfnamefont {M.}~\bibnamefont {Medarde}}, \bibinfo {author} {\bibfnamefont {E.~D.}\ \bibnamefont {Bauer}}, \bibinfo {author} {\bibfnamefont {O.}~\bibnamefont {Zaharko}}, \bibinfo {author} {\bibfnamefont {M.}~\bibnamefont {Kenzelmann}},\ and\ \bibinfo {author} {\bibfnamefont {T.}~\bibnamefont {Park}},\ }\href {https://arxiv.org/abs/2311.13405} {\bibinfo {title} {Triple-sinusoid hedgehog lattice in a centrosymmetric kondo metal}} (\bibinfo {year} {2023}),\ \Eprint {https://arxiv.org/abs/2311.13405} {arXiv:2311.13405 [cond-mat.str-el]}
  \BibitemShut {NoStop}%
\bibitem [{\citenamefont {Ghimire}\ \emph {et~al.}(2015)\citenamefont {Ghimire}, \citenamefont {Ronning}, \citenamefont {Williams}, \citenamefont {Scott}, \citenamefont {Luo}, \citenamefont {Thompson},\ and\ \citenamefont {Bauer}}]{GhimireNJ:JPCM27:2015}%
  \BibitemOpen
  \bibfield  {author} {\bibinfo {author} {\bibfnamefont {N.~J.}\ \bibnamefont {Ghimire}}, \bibinfo {author} {\bibfnamefont {F.}~\bibnamefont {Ronning}}, \bibinfo {author} {\bibfnamefont {D.~J.}\ \bibnamefont {Williams}}, \bibinfo {author} {\bibfnamefont {B.~L.}\ \bibnamefont {Scott}}, \bibinfo {author} {\bibfnamefont {Y.}~\bibnamefont {Luo}}, \bibinfo {author} {\bibfnamefont {J.~D.}\ \bibnamefont {Thompson}},\ and\ \bibinfo {author} {\bibfnamefont {E.~D.}\ \bibnamefont {Bauer}},\ }\bibfield  {title} {\bibinfo {title} {Investigation of the physical properties of the tetragonal {Ce}${M}${Al}$_4${Si}$_2$ (${M}$ = {Rh}, {Ir}, {Pt}) compounds},\ }\href@noop {} {\bibfield  {journal} {\bibinfo  {journal} {J. Phys. :Condens. Matter}\ }\textbf {\bibinfo {volume} {27}},\ \bibinfo {pages} {025601} (\bibinfo {year} {2015})}\BibitemShut {NoStop}%
\bibitem [{\citenamefont {Maurya}\ \emph {et~al.}(2016)\citenamefont {Maurya}, \citenamefont {Kulkarni}, \citenamefont {Thamizhavel}, \citenamefont {Paudyal},\ and\ \citenamefont {Dhar}}]{Maurya2016Kondo-Lattice-a}%
  \BibitemOpen
  \bibfield  {author} {\bibinfo {author} {\bibfnamefont {A.}~\bibnamefont {Maurya}}, \bibinfo {author} {\bibfnamefont {R.}~\bibnamefont {Kulkarni}}, \bibinfo {author} {\bibfnamefont {A.}~\bibnamefont {Thamizhavel}}, \bibinfo {author} {\bibfnamefont {D.}~\bibnamefont {Paudyal}},\ and\ \bibinfo {author} {\bibfnamefont {S.~K.}\ \bibnamefont {Dhar}},\ }\bibfield  {title} {\bibinfo {title} {Kondo lattice and antiferromagnetic behavior in quaternary {Ce}{$T$}{Al}$_4${Si}$_2$ ({$T$} = {Rh}, {Ir}) single crystals},\ }\href {https://doi.org/10.7566/JPSJ.85.034720} {\bibfield  {journal} {\bibinfo  {journal} {Journal of the Physical Society of Japan}\ }\textbf {\bibinfo {volume} {85}},\ \bibinfo {pages} {034720} (\bibinfo {year} {2016})}\BibitemShut {NoStop}%
\bibitem [{\citenamefont {Sakai}\ \emph {et~al.}(2016)\citenamefont {Sakai}, \citenamefont {Hattori}, \citenamefont {Tokunaga}, \citenamefont {Kambe}, \citenamefont {Ghimire}, \citenamefont {Ronning}, \citenamefont {Bauer},\ and\ \citenamefont {Thompson}}]{SakaiH:PRB93:2016}%
  \BibitemOpen
  \bibfield  {author} {\bibinfo {author} {\bibfnamefont {H.}~\bibnamefont {Sakai}}, \bibinfo {author} {\bibfnamefont {T.}~\bibnamefont {Hattori}}, \bibinfo {author} {\bibfnamefont {Y.}~\bibnamefont {Tokunaga}}, \bibinfo {author} {\bibfnamefont {S.}~\bibnamefont {Kambe}}, \bibinfo {author} {\bibfnamefont {N.~J.}\ \bibnamefont {Ghimire}}, \bibinfo {author} {\bibfnamefont {F.}~\bibnamefont {Ronning}}, \bibinfo {author} {\bibfnamefont {E.~D.}\ \bibnamefont {Bauer}},\ and\ \bibinfo {author} {\bibfnamefont {J.~D.}\ \bibnamefont {Thompson}},\ }\bibfield  {title} {\bibinfo {title} {Incommensurate to commensurate antiferromagnetism in {CeRhAl}$_4${Si}$_2$: An $^{27}${Al} {NMR} study},\ }\href@noop {} {\bibfield  {journal} {\bibinfo  {journal} {Phys. Rev. B}\ }\textbf {\bibinfo {volume} {93}},\ \bibinfo {pages} {014402} (\bibinfo {year} {2016})}\BibitemShut {NoStop}%
\bibitem [{\citenamefont {Pyykk{\"o}}(2008)}]{Pyykko2008Year-2008-nucle}%
  \BibitemOpen
  \bibfield  {author} {\bibinfo {author} {\bibfnamefont {P.}~\bibnamefont {Pyykk{\"o}}},\ }\bibfield  {title} {\bibinfo {title} {Year-2008 nuclear quadrupole moments},\ }\href {https://doi.org/10.1080/00268970802018367} {\bibfield  {journal} {\bibinfo  {journal} {Molecular Physics}\ }\textbf {\bibinfo {volume} {106}},\ \bibinfo {pages} {1965} (\bibinfo {year} {2008})}\BibitemShut {NoStop}%
\bibitem [{\citenamefont {Blaha}\ \emph {et~al.}(1992)\citenamefont {Blaha}, \citenamefont {Singh}, \citenamefont {Sorantin},\ and\ \citenamefont {Schwarz}}]{Blaha_rutile_EFGcalc}%
  \BibitemOpen
  \bibfield  {author} {\bibinfo {author} {\bibfnamefont {P.}~\bibnamefont {Blaha}}, \bibinfo {author} {\bibfnamefont {D.~J.}\ \bibnamefont {Singh}}, \bibinfo {author} {\bibfnamefont {P.~I.}\ \bibnamefont {Sorantin}},\ and\ \bibinfo {author} {\bibfnamefont {K.}~\bibnamefont {Schwarz}},\ }\bibfield  {title} {\bibinfo {title} {Electric-field-gradient calculations for systems with large extended-core-state contributions},\ }\href@noop {} {\bibfield  {journal} {\bibinfo  {journal} {Phys. Rev. B}\ }\textbf {\bibinfo {volume} {46}},\ \bibinfo {pages} {1321} (\bibinfo {year} {1992})}\BibitemShut {NoStop}%
\bibitem [{\citenamefont {Matsumura}\ \emph {et~al.}(2009)\citenamefont {Matsumura}, \citenamefont {Kawamura}, \citenamefont {Edamoto}, \citenamefont {Takesaka}, \citenamefont {Kato}, \citenamefont {Nishioka}, \citenamefont {Tokunaga}, \citenamefont {Kambe},\ and\ \citenamefont {Yasuoka}}]{Matsumura2009Novel-Phase-Tra}%
  \BibitemOpen
  \bibfield  {author} {\bibinfo {author} {\bibfnamefont {M.}~\bibnamefont {Matsumura}}, \bibinfo {author} {\bibfnamefont {Y.}~\bibnamefont {Kawamura}}, \bibinfo {author} {\bibfnamefont {S.}~\bibnamefont {Edamoto}}, \bibinfo {author} {\bibfnamefont {T.}~\bibnamefont {Takesaka}}, \bibinfo {author} {\bibfnamefont {H.}~\bibnamefont {Kato}}, \bibinfo {author} {\bibfnamefont {T.}~\bibnamefont {Nishioka}}, \bibinfo {author} {\bibfnamefont {Y.}~\bibnamefont {Tokunaga}}, \bibinfo {author} {\bibfnamefont {S.}~\bibnamefont {Kambe}},\ and\ \bibinfo {author} {\bibfnamefont {H.}~\bibnamefont {Yasuoka}},\ }\bibfield  {title} {\bibinfo {title} {Novel phase transition in {CeRu}$_2${Al}$_{10}$ probed by $^{27}${Al}-{NQR/NMR} --no evidence of magnetic ordering--},\ }\href {https://doi.org/10.1143/JPSJ.78.123713} {\bibfield  {journal} {\bibinfo  {journal} {Journal of the Physical Society of Japan}\ }\textbf {\bibinfo {volume} {78}},\ \bibinfo {pages} {123713} (\bibinfo {year} {2009})}\BibitemShut {NoStop}%
\bibitem [{\citenamefont {Matsuno}\ \emph {et~al.}(2012)\citenamefont {Matsuno}, \citenamefont {Nohara}, \citenamefont {Kotegawa}, \citenamefont {Matsuoka}, \citenamefont {Tomiyama}, \citenamefont {Sugawara}, \citenamefont {Harima},\ and\ \citenamefont {Tou}}]{MatsunoH:JPSJ81:2012}%
  \BibitemOpen
  \bibfield  {author} {\bibinfo {author} {\bibfnamefont {H.}~\bibnamefont {Matsuno}}, \bibinfo {author} {\bibfnamefont {H.}~\bibnamefont {Nohara}}, \bibinfo {author} {\bibfnamefont {H.}~\bibnamefont {Kotegawa}}, \bibinfo {author} {\bibfnamefont {E.}~\bibnamefont {Matsuoka}}, \bibinfo {author} {\bibfnamefont {Y.}~\bibnamefont {Tomiyama}}, \bibinfo {author} {\bibfnamefont {H.}~\bibnamefont {Sugawara}}, \bibinfo {author} {\bibfnamefont {H.}~\bibnamefont {Harima}},\ and\ \bibinfo {author} {\bibfnamefont {H.}~\bibnamefont {Tou}},\ }\bibfield  {title} {\bibinfo {title} {Ising-type magnetic anisotropy derived by {$\Gamma_7^{(2)}$} crystal electric field ground state in tetragonal {CeRu}$_2${Al}$_2${B}: $^{11}${B} and $^{27}${Al} {NMR} studies},\ }\href@noop {} {\bibfield  {journal} {\bibinfo  {journal} {J. Phys. Soc. Jpn.}\ }\textbf {\bibinfo {volume} {81}},\ \bibinfo {pages} {073705} (\bibinfo {year} {2012})}\BibitemShut {NoStop}%
\bibitem [{\citenamefont {Sakai}\ \emph {et~al.}(2012)\citenamefont {Sakai}, \citenamefont {Tokunaga}, \citenamefont {Kambe}, \citenamefont {Baumbach}, \citenamefont {Ronning}, \citenamefont {Bauer},\ and\ \citenamefont {Thompson}}]{SakaiH:PRB86:2012}%
  \BibitemOpen
  \bibfield  {author} {\bibinfo {author} {\bibfnamefont {H.}~\bibnamefont {Sakai}}, \bibinfo {author} {\bibfnamefont {Y.}~\bibnamefont {Tokunaga}}, \bibinfo {author} {\bibfnamefont {S.}~\bibnamefont {Kambe}}, \bibinfo {author} {\bibfnamefont {R.~E.}\ \bibnamefont {Baumbach}}, \bibinfo {author} {\bibfnamefont {F.}~\bibnamefont {Ronning}}, \bibinfo {author} {\bibfnamefont {E.~D.}\ \bibnamefont {Bauer}},\ and\ \bibinfo {author} {\bibfnamefont {J.~D.}\ \bibnamefont {Thompson}},\ }\bibfield  {title} {\bibinfo {title} {{NMR} study for {4f-localized} ferromagnet {CeRu}$_2${Ga}$_2${B}},\ }\href@noop {} {\bibfield  {journal} {\bibinfo  {journal} {Phys. Rev. B}\ }\textbf {\bibinfo {volume} {86}},\ \bibinfo {pages} {094402} (\bibinfo {year} {2012})}\BibitemShut {NoStop}%
\bibitem [{\citenamefont {Kobayashi}\ \emph {et~al.}(2013)\citenamefont {Kobayashi}, \citenamefont {Fujiwara}, \citenamefont {Takeda}, \citenamefont {Harima}, \citenamefont {Ikeda}, \citenamefont {Adachi}, \citenamefont {Ohishi}, \citenamefont {Geibel},\ and\ \citenamefont {Steglich}}]{KobayashiTC:JPSJ82:2013}%
  \BibitemOpen
  \bibfield  {author} {\bibinfo {author} {\bibfnamefont {T.~C.}\ \bibnamefont {Kobayashi}}, \bibinfo {author} {\bibfnamefont {K.}~\bibnamefont {Fujiwara}}, \bibinfo {author} {\bibfnamefont {K.}~\bibnamefont {Takeda}}, \bibinfo {author} {\bibfnamefont {H.}~\bibnamefont {Harima}}, \bibinfo {author} {\bibfnamefont {Y.}~\bibnamefont {Ikeda}}, \bibinfo {author} {\bibfnamefont {T.}~\bibnamefont {Adachi}}, \bibinfo {author} {\bibfnamefont {Y.}~\bibnamefont {Ohishi}}, \bibinfo {author} {\bibfnamefont {C.}~\bibnamefont {Geibel}},\ and\ \bibinfo {author} {\bibfnamefont {F.}~\bibnamefont {Steglich}},\ }\bibfield  {title} {\bibinfo {title} {Valence crossover of {Ce} ions in {CeCu}$_2${Si}$_2$ under high pressure --pressure dependence of the unit cell volume and the {NQR} frequency--},\ }\href@noop {} {\bibfield  {journal} {\bibinfo  {journal} {J. Phys. Soc. Jpn.}\ }\textbf {\bibinfo {volume} {82}},\ \bibinfo {pages} {114701} (\bibinfo {year} {2013})}\BibitemShut {NoStop}%
\bibitem [{\citenamefont {Moriya}(1963)}]{Moriya1963The-Effect-of-E}%
  \BibitemOpen
  \bibfield  {author} {\bibinfo {author} {\bibfnamefont {T.}~\bibnamefont {Moriya}},\ }\bibfield  {title} {\bibinfo {title} {The effect of electron-electron interaction on the nuclear spin relaxation in metals},\ }\href@noop {} {\bibfield  {journal} {\bibinfo  {journal} {J. Phys. Soc. Jpn.}\ }\textbf {\bibinfo {volume} {18}},\ \bibinfo {pages} {516} (\bibinfo {year} {1963})}\BibitemShut {NoStop}%
\bibitem [{\citenamefont {Moriya}(1956)}]{Moriya_LocalMomentLimit}%
  \BibitemOpen
  \bibfield  {author} {\bibinfo {author} {\bibfnamefont {T.}~\bibnamefont {Moriya}},\ }\bibfield  {title} {\bibinfo {title} {Nuclear magnetic relaxation in antiferromagnetics, {I}, {II}},\ }\href@noop {} {\bibfield  {journal} {\bibinfo  {journal} {Prog. Theor. Phys.}\ }\textbf {\bibinfo {volume} {16}},\ \bibinfo {pages} {23,641} (\bibinfo {year} {1956})}\BibitemShut {NoStop}%
\bibitem [{\citenamefont {Moriya}\ and\ \citenamefont {Takimoto}(1995)}]{MoriyaT:JPSJ64:1995}%
  \BibitemOpen
  \bibfield  {author} {\bibinfo {author} {\bibfnamefont {T.}~\bibnamefont {Moriya}}\ and\ \bibinfo {author} {\bibfnamefont {T.}~\bibnamefont {Takimoto}},\ }\bibfield  {title} {\bibinfo {title} {Anomalous properties around magnetic instability in heavy electron systems},\ }\href@noop {} {\bibfield  {journal} {\bibinfo  {journal} {J. Phys. Soc. Jpn.}\ }\textbf {\bibinfo {volume} {64}},\ \bibinfo {pages} {960} (\bibinfo {year} {1995})}\BibitemShut {NoStop}%
\bibitem [{\citenamefont {Kambe}\ \emph {et~al.}(2007)\citenamefont {Kambe}, \citenamefont {Sakai}, \citenamefont {Tokunaga}, \citenamefont {Fujimoto}, \citenamefont {Walstedt}, \citenamefont {Ikeda}, \citenamefont {Aoki}, \citenamefont {Homma}, \citenamefont {Haga}, \citenamefont {Shiokawa},\ and\ \citenamefont {\={O}nuki}}]{Kambe_115T1anisotropy}%
  \BibitemOpen
  \bibfield  {author} {\bibinfo {author} {\bibfnamefont {S.}~\bibnamefont {Kambe}}, \bibinfo {author} {\bibfnamefont {H.}~\bibnamefont {Sakai}}, \bibinfo {author} {\bibfnamefont {Y.}~\bibnamefont {Tokunaga}}, \bibinfo {author} {\bibfnamefont {T.}~\bibnamefont {Fujimoto}}, \bibinfo {author} {\bibfnamefont {R.~E.}\ \bibnamefont {Walstedt}}, \bibinfo {author} {\bibfnamefont {S.}~\bibnamefont {Ikeda}}, \bibinfo {author} {\bibfnamefont {D.}~\bibnamefont {Aoki}}, \bibinfo {author} {\bibfnamefont {Y.}~\bibnamefont {Homma}}, \bibinfo {author} {\bibfnamefont {Y.}~\bibnamefont {Haga}}, \bibinfo {author} {\bibfnamefont {Y.}~\bibnamefont {Shiokawa}},\ and\ \bibinfo {author} {\bibfnamefont {Y.}~\bibnamefont {\={O}nuki}},\ }\bibfield  {title} {\bibinfo {title} {Favorable magnetic fluctuation anisotropy for unconventional superconductivity in $f$-electron systems},\ }\href@noop {} {\bibfield  {journal} {\bibinfo  {journal} {Phys. Rev. B}\ }\textbf {\bibinfo {volume} {75}},\ \bibinfo {pages} {140509(R) (4)} (\bibinfo {year}
  {2007})}\BibitemShut {NoStop}%
\bibitem [{\citenamefont {Sakai}\ \emph {et~al.}(2010)\citenamefont {Sakai}, \citenamefont {Baek}, \citenamefont {Brown}, \citenamefont {Ronning}, \citenamefont {Bauer},\ and\ \citenamefont {Thompson}}]{SakaiH:PRB82:2010}%
  \BibitemOpen
  \bibfield  {author} {\bibinfo {author} {\bibfnamefont {H.}~\bibnamefont {Sakai}}, \bibinfo {author} {\bibfnamefont {S.-H.}\ \bibnamefont {Baek}}, \bibinfo {author} {\bibfnamefont {S.~E.}\ \bibnamefont {Brown}}, \bibinfo {author} {\bibfnamefont {F.}~\bibnamefont {Ronning}}, \bibinfo {author} {\bibfnamefont {E.~D.}\ \bibnamefont {Bauer}},\ and\ \bibinfo {author} {\bibfnamefont {J.~D.}\ \bibnamefont {Thompson}},\ }\bibfield  {title} {\bibinfo {title} {$^{59}${Co} {NMR} shift anomalies and spin dynamics in the normal state of superconducting {CeCoIn}$_5$: Verification of two-dimensional antiferromagnetic spin fluctuations},\ }\href@noop {} {\bibfield  {journal} {\bibinfo  {journal} {Phys. Rev. B}\ }\textbf {\bibinfo {volume} {82}},\ \bibinfo {pages} {020501(R)} (\bibinfo {year} {2010})}\BibitemShut {NoStop}%
\bibitem [{\citenamefont {Baek}\ \emph {et~al.}(2010)\citenamefont {Baek}, \citenamefont {Sakai}, \citenamefont {Bauer}, \citenamefont {Mitchell}, \citenamefont {Kennison}, \citenamefont {Ronning},\ and\ \citenamefont {Thompson}}]{BaekSH:PRL105:2010}%
  \BibitemOpen
  \bibfield  {author} {\bibinfo {author} {\bibfnamefont {S.-H.}\ \bibnamefont {Baek}}, \bibinfo {author} {\bibfnamefont {H.}~\bibnamefont {Sakai}}, \bibinfo {author} {\bibfnamefont {E.~D.}\ \bibnamefont {Bauer}}, \bibinfo {author} {\bibfnamefont {J.~N.}\ \bibnamefont {Mitchell}}, \bibinfo {author} {\bibfnamefont {J.~A.}\ \bibnamefont {Kennison}}, \bibinfo {author} {\bibfnamefont {F.}~\bibnamefont {Ronning}},\ and\ \bibinfo {author} {\bibfnamefont {J.~D.}\ \bibnamefont {Thompson}},\ }\bibfield  {title} {\bibinfo {title} {Anisotropic spin fluctuations and superconductivity in ``115'' heavy fermion compounds: $^{59}${Co} {NMR} study in {PuCoGa}$_5$},\ }\href@noop {} {\bibfield  {journal} {\bibinfo  {journal} {Phys. Rev. Lett.}\ }\textbf {\bibinfo {volume} {105}},\ \bibinfo {pages} {217002} (\bibinfo {year} {2010})}\BibitemShut {NoStop}%
\bibitem [{\citenamefont {Shin~{et al.}}()}]{Shin_H-Tdiagram_preparation}%
  \BibitemOpen
  \bibfield  {author} {\bibinfo {author} {\bibfnamefont {S.}~\bibnamefont {Shin~{et al.}}},\ }\bibfield  {title} {\bibinfo {title} {{$H$}--{$T$} phase diagram of {CePtAl}$_4${Ge}$_2$ for ${H \parallel c}$},\ }\bibinfo {note} {in preparation}\BibitemShut {NoStop}%
\bibitem [{\citenamefont {Takeda}\ \emph {et~al.}(2024)\citenamefont {Takeda}, \citenamefont {Ishikawa}, \citenamefont {Takigawa}, \citenamefont {Yamashita}, \citenamefont {Fujima},\ and\ \citenamefont {Arima}}]{Takeda2024Magnetic-struct}%
  \BibitemOpen
  \bibfield  {author} {\bibinfo {author} {\bibfnamefont {H.}~\bibnamefont {Takeda}}, \bibinfo {author} {\bibfnamefont {M.}~\bibnamefont {Ishikawa}}, \bibinfo {author} {\bibfnamefont {M.}~\bibnamefont {Takigawa}}, \bibinfo {author} {\bibfnamefont {M.}~\bibnamefont {Yamashita}}, \bibinfo {author} {\bibfnamefont {Y.}~\bibnamefont {Fujima}},\ and\ \bibinfo {author} {\bibfnamefont {T.-h.}\ \bibnamefont {Arima}},\ }\bibfield  {title} {\bibinfo {title} {Magnetic structure of polar magnet {GaV}$_{4}${Se}$_{8}$ with {N\'eel}-type skyrmion lattice probed by $^{51}${V} {NMR}},\ }\href {https://doi.org/10.1103/PhysRevB.110.224430} {\bibfield  {journal} {\bibinfo  {journal} {Phys. Rev. B}\ }\textbf {\bibinfo {volume} {110}},\ \bibinfo {pages} {224430} (\bibinfo {year} {2024})}\BibitemShut {NoStop}%
\bibitem [{\citenamefont {Park}\ \emph {et~al.}(2025)\citenamefont {Park}, \citenamefont {Sakai}, \citenamefont {Hosoi}, \citenamefont {Thomas}, \citenamefont {Kambe}, \citenamefont {Tokunaga}, \citenamefont {Dioguardi}, \citenamefont {Thompson}, \citenamefont {Ronning}, \citenamefont {Kimata}, \citenamefont {Furukawa}, \citenamefont {Sasaki}, \citenamefont {Bauer},\ and\ \citenamefont {Hirata}}]{Park2025Investigation-o}%
  \BibitemOpen
  \bibfield  {author} {\bibinfo {author} {\bibfnamefont {S.}~\bibnamefont {Park}}, \bibinfo {author} {\bibfnamefont {H.}~\bibnamefont {Sakai}}, \bibinfo {author} {\bibfnamefont {S.}~\bibnamefont {Hosoi}}, \bibinfo {author} {\bibfnamefont {S.~M.}\ \bibnamefont {Thomas}}, \bibinfo {author} {\bibfnamefont {S.}~\bibnamefont {Kambe}}, \bibinfo {author} {\bibfnamefont {Y.}~\bibnamefont {Tokunaga}}, \bibinfo {author} {\bibfnamefont {A.~P.}\ \bibnamefont {Dioguardi}}, \bibinfo {author} {\bibfnamefont {J.~D.}\ \bibnamefont {Thompson}}, \bibinfo {author} {\bibfnamefont {F.}~\bibnamefont {Ronning}}, \bibinfo {author} {\bibfnamefont {M.}~\bibnamefont {Kimata}}, \bibinfo {author} {\bibfnamefont {T.}~\bibnamefont {Furukawa}}, \bibinfo {author} {\bibfnamefont {T.}~\bibnamefont {Sasaki}}, \bibinfo {author} {\bibfnamefont {E.~D.}\ \bibnamefont {Bauer}},\ and\ \bibinfo {author} {\bibfnamefont {M.}~\bibnamefont {Hirata}},\ }\href {https://arxiv.org/abs/2506.04563} {\bibinfo {title} {Investigation of the paramagnetic state of
  the kagome kondo lattice compound {YbV}$_6${Sn}$_6$: a $^{51}${V} nuclear magnetic resonance study}},\ \Eprint {https://arxiv.org/abs/2506.04563} {arXiv:2506.04563} \BibitemShut {NoStop}%
\end{thebibliography}

%

\end{document}